\documentclass[a4paper,11pt]{article}
\pdfoutput=1 

\usepackage{jcappub}

\makeatletter
\gdef\@fpheader{}
\g@addto@macro\bfseries{\boldmath}
\makeatother

\usepackage{amssymb,amsmath,amsfonts,amsbsy,graphicx}
\usepackage{color}
\usepackage{psfrag}
\usepackage{tikz}
\usepackage{xspace}
\usepackage[normalem]{ulem} 
\usepackage{amssymb}
\usepackage{amsmath}
\usepackage{amsbsy} 
\usepackage{bm}
\usepackage{graphicx}
\usepackage{color}
\usepackage{subfig}
\usepackage{caption} 
\usepackage{graphicx}
\usepackage{url}

\usepackage{tikz}
\usetikzlibrary{positioning}
\usepackage{siunitx}
\usetikzlibrary{intersections}
\usetikzlibrary{decorations}
\usetikzlibrary{decorations.pathmorphing}
\usetikzlibrary{decorations.pathreplacing}
\usetikzlibrary{decorations.shapes}
\usetikzlibrary{decorations.text}
\usetikzlibrary{decorations.markings}
\usetikzlibrary{decorations.fractals}
\usetikzlibrary{decorations.footprints}

\usepackage{stackrel}

\newcommand{\ie}{\textsl{i.e.~}}

\newcommand{\eg}{\textsl{e.g.~}}

\newcommand{\etc}{\textsl{etc.~}}

\newcommand{\dd}{\mathrm{d}}
\newcommand{\ee}{e}

\newcommand{\sss}[1]{{\scriptscriptstyle{#1}}}

\newcommand{\uPl}{\mathrm{Pl}}
\newcommand{\uin}{\mathrm{in}}

\newcommand{\uend}{\mathrm{end}}

\newcommand{\ucl}{\mathrm{cl}}
\newcommand{\ueq}{\mathrm{eq}}

\newcommand{\uc}{\mathrm{c}}

\newcommand{\usssPl}{\sss{\uPl}}

\newcommand{\calP}{\mathcal{P}}

\newcommand{\MeV}{\mathrm{MeV}}
\newcommand{\GeV}{\mathrm{GeV}}

\newcommand{\Mpc}{\mathrm{Mpc}}

\newcommand{\Mp}{M_\usssPl}

\newcommand{\efolds}{$e$-folds\xspace}

\newcommand{\Nend}{N_\uend}

\newcommand{\beq}{\begin{equation}}
\newcommand{\eeq}{\end{equation}}
\newcommand{\bea}{\begin{equation}\begin{aligned}}
\newcommand{\eea}{\end{aligned}\end{equation}}

\newlength{\wsingfig}
\newlength{\wdblefig}
\newlength{\wquadfig}
\newlength{\wtriplefig}
\newcommand{\Eq}[1]{eq.~(\ref{#1})}
\newcommand{\Eqs}[1]{eqs.~(\ref{#1})}
\newcommand{\Fig}[1]{fig.~{\ref{#1}}}

\newcommand{\Refa}[1]{ref.~{\cite{#1}}}
\newcommand{\Refs}[1]{refs.~{\cite{#1}}}
\newcommand{\Sec}[1]{sec.~\ref{#1}}
\newcommand{\Secs}[1]{secs.~\ref{#1}}
\newcommand{\App}[1]{appendix~\ref{#1}}

\newcommand{\Hini}{{H_{\uin}}}

\newcommand{\Hc}{{H_{\uc}}}
\newcommand{\Hend}{{H_{\uend}}}

\newcommand{\Nt}{{N_{\mathrm{t}}}}
\newcommand{\Neq}{{N_{\ueq}}}
\newcommand{\NCST}{{N_{\mathrm{CST}}}}

\title{Revisiting the stochastic QCD axion window: departure from equilibrium during inflation}

\author[a]{Vadim Briaud,}
\author[b,c]{Kenji Kadota,}
\author[d,e]{Shinji Mukohyama,}
\author[f]{Alireza Talebian,}
\author[a]{Vincent Vennin}

\affiliation[a]{Laboratoire de Physique de l'École Normale Supérieure, ENS, CNRS, Université PSL, Sorbonne Université, Université Paris Cité, F-75005 Paris, France}
\affiliation[b]{School of Fundamental Physics and Mathematical Sciences,
  Hangzhou Institute for Advanced Study, University of Chinese Academy of Sciences (HIAS-UCAS), Hangzhou 310024, China}
 \affiliation[c]{International Centre for Theoretical Physics Asia-Pacific (ICTP-AP), Beijing/Hangzhou, China
  }
\affiliation[d]{Center for Gravitational Physics and Quantum Information, Yukawa Institute for Theoretical Physics, Kyoto University, 606-8502, Kyoto, Japan}
\affiliation[e]{Kavli Institute for the Physics and Mathematics of the Universe (WPI), The University of Tokyo Institutes for Advanced Study (UTIAS), The University of Tokyo, Kashiwa, Chiba 277-8583, Japan}
\affiliation[f]{School of Astronomy, Institute for Research in Fundamental Sciences (IPM) P. O. Box 19395-5531, Tehran, Iran}
\emailAdd{vadim.briaud@ens.fr}
\emailAdd{kadota@ucas.ac.cn}
\emailAdd{shinji.mukohyama@yukawa.kyoto-u.ac.jp}
\emailAdd{talebian@ipm.ir}
\emailAdd{vincent.vennin@ens.fr}

\date{today}

\begin{document}

\begin{flushright} {YITP-23-161\\IPMU23-0047}  \end{flushright}
\vspace{-1.4cm}

\sloppy

\abstract{If dark matter is made of QCD axions, its abundance is determined by the vacuum expectation value acquired by the axion field during inflation. The axion is usually assumed to follow the equilibrium distribution arising from quantum diffusion during inflation. This leads to the so-called stochastic window under which the QCD axion can make up all the dark matter. It is characterised by $10^{10.4}\GeV\leq f\leq 10^{17.2}\GeV$ and  $\Hend>10^{-2.2}\GeV$, where $f$ is the axion decay constant and $\Hend$ is the Hubble expansion rate at the end of inflation. However, in realistic inflationary potentials, we show that the axion never reaches the equilibrium distribution at the end of inflation. This is because the relaxation time of the axion is much larger than the typical time scale over which $H$ varies during inflation. As a consequence, the axion acquires a quasi-flat distribution as long as it remains light during inflation. This leads us to reassessing the stochastic axion window, and we find that $ 10^{10.3}\GeV\leq f\leq 10^{14.1}\GeV$ and $\Hend>10^{-13.8}\GeV$.}
 

\maketitle

\section{Introduction}
\label{sec:intro}

The identity of dark matter remains an open question, and a wide range of possibilities has been explored. Examples include axion-like dark matter with a mass of order \(10^{-21} \, \text{eV}\), weakly interacting massive particles with weak-scale mass (\(\sim 10^2 \, \text{GeV}\)), and dark matter with mass beyond the Planck scale, such as primordial black holes \cite{Jungman:1995df,Bertone:2004pz,Peccei:1977hh,Weinberg:1977ma,Wilczek:1977pj,Preskill:1982cy,Abbott:1982af,Dine:1982ah, Sikivie:1983ip,Choi:2020rgn,Carr:2020xqk,Green:2020jor, Carr:2020gox,Hui:2016ltb, Hui:2021tkt,Firouzjahi:2021lov,Safarzadeh:2019sre,Talebian:2022jkb,Talebian:2022cwk}. While the mass of dark matter can vary widely, light dark-matter candidates are naturally expected from the likely symmetry breakings that occur in the early Universe, such as pseudo-Nambu-Goldstone bosons originating from the spontaneous breaking of a continuous global symmetry. They are worth exploring for the role they may play in various cosmological mechanisms and for the observational imprints they may leave~\cite{Lyth:2001nq, Bartolo:2002vf, Ringeval:2010hf, Herranen:2014cua, Kearney:2015vba, Vennin:2015vfa, Hardwick:2017qcw, Enqvist:2017kzh, Torrado:2017qtr, Takahashi:2019pqf}. The axion is a well-known example, arising from the spontaneous breaking of Peccei-Quinn (PQ) symmetry in the early Universe, and was originally introduced to solve the QCD strong-CP problem \cite{Peccei:1977hh,Weinberg:1977ma,Wilczek:1977pj,Preskill:1982cy,Abbott:1982af,Dine:1982ah,Sikivie:1983ip,Choi:2020rgn, Gorghetto:2023vqu}.

The cosmological role played by such light particles depends on the vacuum expectation value they acquire during inflation. In the case of dark-matter axions for instance, the misalignment angle at the onset of axion oscillations sets the eventual dark-matter abundance. The goal of this paper is to calculate the probability distribution of the QCD axion misalignment angle, and study its dependence on the axion decay constant and on the energy scale of inflation. 

It has first been assumed that \cite{Preskill:1982cy,Dine:1982ah,Abbott:1982af}, in the absence of fine tuning, the initial misalignment angle $\theta_\uend$ (where ``end'' denotes the end of inflation, hence the ``initial'' condition for the subsequent evolution) should be of order unity, due to the $U(1)_{\text{PQ}}$ symmetry. Hence, a flat probability distribution for $\theta_\uend$ is frequently adopted. For the axion decay constant $f$, this leads to the so-called \textit{classical} axion window 
$10^8 \GeV\lesssim f\lesssim 10^{12}\GeV$. The upper bound comes from the requirement for the axion not to exceed the observed dark matter abundance and the lower bound comes from the supernova neutrino and stellar cooling constraints~\cite{Turner:1989vc,Kaplan:1985dv,Dine:1982ah,Abbott:1982af,Mayle:1987as,Raffelt:1987yt,Turner:1987by, Preskill:1982cy,Raffelt:2006cw}.
 
If $\theta_\uend\ll 1$, larger values of $f$ (hence smaller values of the axion mass $m\propto 1/f$) are allowed, but this requires fine tuning.

Recently however, it has been pointed out \cite{Graham:2018jyp, Takahashi:2018tdu} that small values of $\theta_\uend$ (hence  values of $f$ larger than $10^{12}\GeV$) are to be expected if one considers the dynamics of the axion field during a phase of low-energy inflation. During inflation, vacuum quantum fluctuations of the axion field are amplified by gravitational instability, and stretched beyond the Hubble radius. This leads to a classical field value for the axion at large scales that arises from the stochastic accumulation of quantum fluctuations throughout inflation, and that can be computed within the formalism of stochastic inflation~\cite{Starobinsky:1982ee, Starobinsky:1986fx}. If the Hubble expansion rate $H=\Hend$ is constant during inflation, the axion acquires an equilibrium distribution, which is almost flat if $H_\uend$ is larger than the QCD scale $\Lambda\sim 100\MeV$, but is peaked otherwise. Indeed, if $\Hend <\Lambda$, the QCD phase transition occurs before the end of inflation, hence the axion acquires a mass and it rolls down its potential until classical rolling and quantum diffusion balance out. The conclusion is that, when $\Hend\lesssim 100\MeV$, $f$ needs to be larger than $10^{12}\GeV$ for the axion to constitute all of dark matter, and when $\Hend<1\MeV$ this even requires $f>\Mp$, where $\Mp\simeq 2.4\times 10^{18}\GeV$ is the reduced Planck mass. This leads to $10^{10.4}\GeV\leq f\leq 10^{17.2}\GeV$ and 
$\Hend>10^{-2.2}\GeV$, which is refereed to as the \textit{stochastic} axion window.     

The assumption that $H$ is constant during inflation seems well justified given the Cosmic Microwave Background (CMB) temperature and polarisation measurements~\cite{Planck:2018jri}, which are compatible with slow-roll models of inflation where $H$ varies slowly. However, as pointed out in \Refa{Hardwick:2017fjo}, and as we will further elaborate on in this work, the typical e-folding time scale required for the axion to reach the equilibrium is of the order $N\sim H^2/m^2$ (where $N=\ln a $ is the number of \efolds and $a$ is the scale factor), which is of order $N\sim 10^{20\sim 30}$ for the parameters of interest. Over such time scales $H$ cannot be considered as constant in most slow-roll models of inflation, and if it can, we will show that inflation has to proceed in the so-called ``eternal'' regime for dark matter to be produced with the right abundance. The stochastic axion window thus needs to be reassessed. 

In this work we will show that, for realistic inflationary potentials, the axion never reaches the equilibrium distribution at the end of inflation. Instead, we find that for $\Hend>10^{-13.8}\GeV$, the axion remains light until inflation ends and $\theta_\uend$ acquires a flat distribution. For smaller values of $\Hend$, the distribution of $\theta_\uend$ is more peaked, hence larger values of $f$ are allowed. Requiring that $f$ be sub-Planckian leads to the constraint $\Hend>10^{-20}\GeV$, which corresponds to the scale of big-bang nucleosynthesis anyway. Therefore this leaves the scale of inflation unconstrained, in stark contrast to the lower bound $H>10^{-2.2}\GeV$ obtained assuming stochastic equilibrium. We are thus led to a drastic reappraisal of the stochastic axion window.

This article is organised as follows. In \Sec{sec:stochastic:axion}, we introduce the stochastic formalism to describe the dynamics of the axion field during inflation, and we show why the equilibrium assumption is in general not valid. In \Sec{sec:distribution end inflation}, we compute the probability distribution of the misalignment angle at the end of inflation, assuming inflation is realised in a potential with quadratic corrections at large-field values. We show that our results are robust under considering other large-field completions, under introducing a prolonged phase of plateau inflation towards the end of inflation, and under changes of initial conditions provided inflation starts above a critical scale that we determine. In \Sec{sec:Stochastic:Window}, we compute the abundance of axion dark matter and we reassess the stochastic axion window. Given that $\theta_\uend$ is a parameter that is stochastic in nature, we make use of a Bayesian framework to derive consistent constraints on $f$ and $\Hend$. We summarise our main results in \Sec{sec:Conclusion}, and we end this article by an appendix where a necessary and sufficient condition for the axion to oscillate during inflation is derived.

\section{The stochastic axion}
\label{sec:stochastic:axion}

\subsection{Stochastic formalism}

The stochastic formalism~\cite{Starobinsky:1982ee, Starobinsky:1986fx} is an effective theory that describes the long-wavelength part of quantum fields on an inflating cosmological background. Here, we consider the axion as a spectator field, hence its fluctuations do not source the background and we treat the inflaton as a classical field without fluctuations.

To obtain this effective theory for the axion, one has to coarse-grain $ \theta $, and its momentum $ \pi $, at the comoving scale $R=(\varepsilon aH)^{-1}$, where $\varepsilon\ll 1$ is the ratio between the Hubble radius and the coarse-graining radius. The infrared (IR) part of the fields is thus defined according to
\bea
\hat{\theta}_{\mathrm{IR}}(\vec{x},N)&=\int\frac{\dd\vec{k}}{(2\pi)^{3/2}} W\left(\frac{k}{\varepsilon a H}\right)\left[ \ee^{-i \vec{k}\cdot\vec{x}}\theta_{k}(N) \hat{a}_{\vec{k}}+\ee^{i \vec{k}\cdot\vec{x}}\theta_{k}^*(N) \hat{a}^\dagger_{\vec{k}}\right]\, ,\\
\hat{\pi}_{\mathrm{IR}}(\vec{x},N)&=\int\frac{\dd\vec{k}}{(2\pi)^{3/2}} W\left(\frac{k}{\varepsilon a H}\right) \left[ \ee^{-i \vec{k}\cdot\vec{x}}\pi_{k}(N) \hat{a}_{\vec{k}}+\ee^{i \vec{k}\cdot\vec{x}}\pi_{k}^*(N) \hat{a}^\dagger_{\vec{k}}\right]\, .
\eea
In this expression, we have introduced $\hat{a}_{\vec{k}}$ and $\hat{a}_{\vec{k}}^\dagger$ which are annihilation and creation operators, hence we are dealing with quantum operators denoted by hats, and $W$ is a window function that selects modes $k$ with wavelength larger than the coarse-graining scale $1/(\varepsilon H)$, \ie $W(x)\simeq 1$ for $x\ll 1$ and $0$ for $x\gg 1$. On super-Hubble scales, gradient terms appearing in the Klein-Gordon equation for $ \theta $ can be neglected and the mode functions $\theta_k$ and $\pi_k$ follow the same equations of motion as the classical homogeneous background. This is the so-called separate-universe approximation~\cite{Salopek:1990jq, Sasaki:1995aw, Wands:2000dp, Pattison:2019hef, Artigas:2021zdk}. By differentiating the above expressions with respect to time, here labeled by the number of \efolds $N=\ln a$, one obtains~\cite{Grain:2017dqa}
\bea 
\label{eq:Langevin:interm}
\frac{\dd\hat{\theta}_{\mathrm{IR}}}{\dd N}&=\hat{\pi}_{\mathrm{IR}}+\hat{\xi}_\theta(N)\, ,\\
\frac{\dd\hat{\pi}_{\mathrm{IR}}}{\dd N}&=-\left(3-\epsilon_1\right)\hat{\pi}_{\mathrm{IR}}- \frac{v_{\mathrm{ax}}'\left(\hat{\theta}_{\mathrm{IR}}\right)}{H^2}+\hat{\xi}_\pi(N)\, ,
\eea
where $ \epsilon_1 = -{\dd \ln H}/{\dd N} $ is the first slow-roll parameter, $v_{\mathrm{ax}}$ is the axion potential and the source functions $\hat{\xi}_\theta$ and $\hat{\xi}_\pi$ read
\bea
\hat{\xi}_\theta(N)&=-\int\frac{\dd\vec{k}}{(2\pi)^{3/2}} \left[ \ee^{-i \vec{k}\cdot\vec{x}}\theta_{k}(N) \hat{a}_{\vec{k}}+\ee^{i \vec{k}\cdot\vec{x}}\theta_{k}^*(N) \hat{a}^\dagger_{\vec{k}}\right]\frac{\dd }{\dd N}W\left(\frac{k}{\varepsilon a H }\right)\, ,\\
\hat{\xi}_\pi(N)&=-\int\frac{\dd\vec{k}}{(2\pi)^{3/2}} \left[ \ee^{-i \vec{k}\cdot\vec{x}}\pi_{k}(N) \hat{a}_{\vec{k}}+\ee^{i \vec{k}\cdot\vec{x}}\pi_{k}^*(N) \hat{a}^\dagger_{\vec{k}}\right]\frac{\dd }{\dd N}W\left(\frac{k}{\varepsilon a H }\right)\, .
\eea

In the effective theory of stochastic dynamics, \Eqs{eq:Langevin:interm} can be replaced with Langevin equations, where quantum operators are replaced with random variables and hats are dropped. In particular, since $\xi_\theta$ and $\xi_\pi$ involve Fourier modes around the coarse-graining scale, where cosmological perturbations still pertain to the perturbative regime, they can be described as Gaussian random variables, with statistical mean and covariance respectively given by the one and two-point vacuum expectation values of $\hat{\xi}_\theta$ and $\hat{\xi}_\pi$.

In this work, we assume the axion to have a potential of the form~\cite{bo2016,Blinov:2019rhb,Borsanyi:2015cka}
\bea
    v_{\mathrm{ax}}\left(\theta\right) = m^2\left(T\right)\left[1-\cos\left(\theta\right)\right]\, ,\quad\text{where}\quad
m\left(T\right)\sim
\begin{cases}
  \frac{\Lambda^2}{f} \left(\frac{T_{\mathrm{QCD}}}{T}\right)^{n} & T > T_{\mathrm{QCD}} \\
   \frac{\Lambda^2}{f} & T\leq T_{\mathrm{QCD}}
  \end{cases}\, ,
  \label{mass temperature}
\eea
where $n$ is a positive constant.
Note that, here, the ``axion'' refers to the angular field $\theta$, which is dimensionless, hence the potential has the dimension of a mass squared. The effective mass of the axion depends on the background temperature, which  we take to be the de Sitter temperature $ T =  {H}/({2\pi}) $ during inflation. For the QCD axion, the phase transition occurs at $T_{\mathrm{QCD}}=150\,\MeV$, and one has $n=4$ and $ \Lambda \simeq 100\, \MeV $. As long as $H\gg m(T)$, the axion follows the slow-roll attractor. In this regime, the phase space becomes effectively one dimensional~\cite{Grain:2017dqa}, and the stochastic dynamics of the axion reduces to
\begin{equation}
\label{Langevin equation in SR}
    \frac{\dd\theta}{\dd N} = -\frac{v_{\mathrm{ax}}'\left(\theta\right)}{3H^2\left(N\right)} + \frac{H\left(N\right)}{2\pi f}\xi(N),
\end{equation}
where $ \xi $ is a white Gaussian noise with unit variance, $\langle \xi(N)\rangle=0$ and $\langle \xi(N) \xi(N')\rangle=\delta(N-N')$, if the window function $W$ is set to a Heaviside function. 

The Langevin equation~\eqref{Langevin equation in SR} can be recast in the form of a Fokker-Planck equation~\cite{Risken:1984book}, which governs the dynamics of the probability distribution function $ P\left(\theta,N\lvert\theta_\uin,N_\uin\right) $ that the axion takes value $ \theta $ at e-folding time $ N $ under the initial condition $ \theta\left(N_\uin\right) = \theta_\uin $. The Fokker-Planck equation takes the form
\bea
\label{Fokker-Planck}
\frac{\partial P}{\partial N} 
= \frac{1}{3H^2(N)}\frac{\partial}{\partial\theta}\left(v_{\mathrm{ax}}' P\right)+ \frac{H^2(N)}{8\pi^2f^2}\frac{\partial^2 P}{\partial\theta^2}
\equiv \mathcal{L}_{\mathrm{FP}}P\, ,
\eea
which defines the Fokker-Planck operator $ \mathcal{L}_{\mathrm{FP}} $.
If the Hubble expansion rate does not depend on time, neither does the axion potential and then the Fokker-Planck equation  admits the stationary solution,
\bea
    \label{stationary solution Fokker-Planck equation}
P_{\mathrm{stat}}(\theta) = \frac{\exp\left[{\frac{8\pi^2m^2f^2}{3H^4}\cos(\theta)}\right]}{2\pi I_0\left(\frac{8\pi^2m^2f^2}{3H^4}\right)}\, .
\eea 
Here, $I_0$ is the modified Bessel function of the first kind,\footnote{When equating the right-hand side of \Eq{Fokker-Planck} to zero, one obtains a second-order differential equation for $P_{\mathrm{stat}}(\theta)$, which thus admits two independent solutions. However, when imposing $2\pi$-periodic boundary conditions, only one survives, and the overall constant in \Eq{stationary solution Fokker-Planck equation} is set such that the distribution is normalised. The same result is obtained if $\theta$ is not a periodic variable, by requiring that the distribution remains finite.} and the $H$-dependence of $m$ through \Eq{mass temperature} with $T=H/(2\pi)$ is left implicit.

\subsection{Beyond the equilibrium distribution}
\label{sec:BeyondEquilibrium}

As already alluded to in the introduction, the stochastic dynamics of the axion depends on the details of the inflationary expansion, which is encoded in the $H(N)$ function. Given that single-scalar-field models of inflation are compatible with cosmological observations, in particular the temperature and polarisation fluctuations of the Cosmic Microwave Background (CMB)~\cite{Planck:2018nkj}, in this work we assume that inflation is driven by a single scalar field $\phi$ with potential energy $V(\phi)$. The models that provide the best agreement with the CMB data are of the plateau type~\cite{Martin:2013tda, Martin:2013nzq}, \ie the $V(\phi)$ function asymptotes to a constant at large-field values. 
In this case, the Hubble expansion rate can be approximated as a constant throughout inflation, and one might expect that the axion reaches the equilibrium distribution~\eqref{stationary solution Fokker-Planck equation}. This is why this distribution is commonly used in the literature to set the initial condition for the axion at the end of inflation \cite{Graham:2018jyp, Takahashi:2018tdu}, where $H$ is replaced by its value at the end of inflation. 

The time required to reach the equilibrium can be estimated as follows. If $H^2\ll m f$, the equilibrium distribution is highly peaked around $\theta\simeq 0$, so one can Taylor expand the potential function $v_{\mathrm{ax}}\simeq m^2 \theta^2/2$ around the origin. In this limit, provided that the initial distribution is Gaussian, the Fokker-Planck equation~\eqref{Fokker-Planck} admits Gaussian solutions 
\bea
\label{eq:Gaussian}
P(\theta,N) = \frac{\ee^{-\frac{\left[\theta-\theta_\ucl(N)\right]^2}{2 \sigma^2(N)}}}{\sqrt{2\pi\sigma^2(N)}}\, ,
\eea 
where $\theta_\ucl(N)$ obeys the Langevin equation~\eqref{Langevin equation in SR} in the absence of noise term, and $\sigma$ obeys $\dd\sigma/\dd N=-m^2\sigma/(3H^2)+H^2/(8\pi^2f^2\sigma)$. These lead to~\cite{Hardwick:2017fjo}
\bea
\label{eq:thetacl:sigma:Gauss:quadInfl}
\theta_\ucl(N) = &\theta_\ucl\left(N_\uin\right)\exp\left[-\frac{m^2}{3H^2}\left(N-N_\uin\right)\right]\, ,\\
\sigma^2(N) =& \frac{3H^4}{8\pi^2 m^2 f^2} + \exp\left[-\frac{2m^2}{3H^2}\left(N-N_\uin\right)\right] \left[\sigma^2\left(N_\uin\right) -\frac{3H^4}{8\pi^2 m^2 f^2} \right]\, .
\eea 
From these expressions, it is clear that the stationary distribution is reached when $N-N_\uin \gg H^2/m^2$, hence $\Neq=H^2/m^2$ is the equilibration e-folding time scale. 

In the opposite regime where $H^2\gg mf$, the equilibrium distribution~\eqref{stationary solution Fokker-Planck equation} is almost flat. This implies that the drift term in the Langevin equation~\eqref{Langevin equation in SR}, induced by the gradient of the potential, can be neglected. In this limit, in the absence of boundary conditions,  the Fokker-Planck equation~\eqref{Fokker-Planck} still admits Gaussian solutions of the form~\eqref{eq:Gaussian}, where $\theta_\ucl(N)=\theta_\uin$ and $\sigma^2(N)=\sigma^2(N_\uin)+H^2(N-N_\uin)/(4\pi^2f^2)$. The standard deviation becomes larger than the period $2\pi$ of the potential after an e-folding time $N_{\ueq}=16\pi^4 f^2/H^2$, which thus corresponds to the equilibration time scale.\footnote{If periodic boundary conditions are added at $\theta=-\pi$ and $\theta=\pi$, the distribution function can still be obtained by the method of images~\cite{Risken:1984book} and one finds
\bea
\label{eq:PGauss:periodic:elliptic}
P(\theta)= \frac{1}{2\pi} \vartheta_3\left[\frac{\theta-\theta_\uin}{2},\ee^{-\frac{\sigma^2(N)}{2}}\right]\, ,
\eea 
where $\vartheta_3$ is the third elliptic theta function. One can check that, when $\sigma\gg 2\pi$, this distribution is quasi flat.\label{PGauss:periodic:elliptic}}
In summary, we find
\bea
\label{eq:Neq}
\Neq=
\begin{cases}
16\pi^4 \frac{f^2}{H^2} \quad & \text{if} \quad H\gg\Lambda\\
\frac{H^2}{m^2} \quad & \text{if} \quad H\ll\Lambda
\end{cases} \, .
\eea

However, by performing observations within our observable horizon, we are only able to probe the last $\sim 60$ \efolds of inflation, and the shape of the inflationary potential outside this window is unknown. In practice, most high-energy constructions that provide plateau potentials are subject to corrections (either radiative corrections, coming from the fact that $\phi$ is a quantum field on a curved background, or corrections arising from the coupling between $\phi$ and other degrees of freedom in the theory, or corrections related to Planck-suppressed higher-order operators, \etc -- for a review, see \Refa{Martin:2013tda}). These corrections usually take a monomial form at large-field values. In other words, in most cases, the potential is of the plateau shape for the last $\sim 100$ \efolds of inflation or so, and monomial before. In the regions of parameter space where $\Neq\gg 100$, the stochastic distribution of the axion field at the end of inflation barely depends on the last $\sim 100$ \efolds of inflation, and rather probes the monomial part of the inflationary potential. 

This is why, in what follows, we will consider the stochastic dynamics of the axion in quadratic inflation where 
\bea
\label{eq:quad:pot:phi}
V\left(\phi\right) = \frac{m^2_\phi}{2} \phi^2\, .
\eea 
The monomial part of the inflationary potential could assume other powers than quadratic (for instance, quartic) and this possibility will be discussed in \Sec{sec:alternative:completions}, but we chose to focus the main analysis on quadratic inflation since corrected mass terms seem ubiquitous in most high-energy constructions. The role of the plateau phase occurring towards the end of inflation will also be quantified in \Sec{sec:plateau}. In the slow-roll regime, \Eq{eq:quad:pot:phi} leads to~\cite{Martin:2013tda}
\begin{equation}
\label{Hubble time}
    H\left(N\right) = \Hend\sqrt{1+2\left(\Nend-N\right)},
\end{equation}
where ``end'' denotes the end of inflation. The duration of inflation starting from $H_\uin$ is thus given by $\Nend-N_\uin=(H_\uin^2/H_\uend^2-1)/2$.

In passing, let us note that if one insisted to use a plateau potential throughout the entire phase of inflation in order for the axion to reach the equilibrium distribution, one would have to start in the regime of so-called ``eternal inflation'' where the curvature perturbation $\zeta$ is large and where the present formalism breaks down. Indeed, in typical plateau models, the $H(N)$ function is of the form~\cite{Roest:2013fha, Martin:2016iqo} $H(N) = \bar{H} \ee^{-\gamma/(N_\uend-N)} $, where $\gamma$ is of order one (for instance $\gamma=\frac{3}{4}$ for Starobinsky or Higgs inflation~\cite{Martin:2013tda}) and depends on the detailed shape of the plateau. For the axion to equilibrate before the end of inflation, inflation needs to last $N_\uend-N_\uin>N_\ueq$, where $N_\ueq$ is given by \Eq{eq:Neq}. Here we have used the fact that $\Lambda$ and $T_{\mathrm{QCD}}$ are of the same order and that $n>0$. This places a lower bound on the amplitude of curvature perturbations at initial time. Indeed, the reduced power spectrum of the curvature perturbation is given by $\calP_\zeta = H^2/( 8\pi^2\Mp^2 \epsilon_1)$ where $\epsilon_1=-\dd\ln H/\dd N $ is the first Hubble-flow parameter, and requiring that $\calP_\zeta$ remains small throughout inflation implies that $f\ll  (\Lambda/H)^2\sqrt{H\Mp}$ when $H\ll\Lambda$ and that $f\ll  \sqrt{H\Mp}$ when $H\gg\Lambda$. Below (see \Fig{fig:EB-2d-1plot}) we show that these conditions are hardly met for parameters leading to the observed abundance of dark matter, hence plateau inflation has to start in the eternal regime for the equilibrium distribution to be reached.

\section{Distribution of the axion at the end of inflation}
\label{sec:distribution end inflation}

In this section, we show how the probability distribution of the axion at the end of inflation can be computed. We focus on scenarios in which the PQ symmetry breaking occurs before or during inflation, which imposes that $H_\uend<f$. Inflation is assumed to start at the scale $ \Hini $, where the axion has initial distribution $P_\uin(\theta)$. As mentioned above, the axion is not expected to reach the equilibrium state~\eqref{stationary solution Fokker-Planck equation} during inflation, hence its final distribution may depend on the initial one, which is why we introduce it explicitly. 

\subsection{The three regimes of the axion dynamics}

The typical e-folding time scale over which $H$ varies is given by $N_H=\epsilon_1^{-1}=\vert \dd\ln H/\dd N\vert^{-1}$, which reduces to $N_H=H^2/H_\uend^2$ in quadratic inflation, see \Eq{Hubble time}. This has to be compared with the equilibration e-folding time scale of the axion, given by \Eq{eq:Neq}. Only when $\Neq\ll N_H$ is the equilibrium distribution tracked adiabatically~\cite{Hardwick:2017fjo}. In this section, we determine when this is the case, and we show that the stochastic dynamics of the axion can be divided into three phases: a diffusion-dominated regime, a frozen regime, and a drift-dominated regime.

\subsubsection{Diffusion-dominated regime}
\label{sec:diffusion:dominated}

At high energy, \ie for large values of $H$, the noise term in the Langevin equation~\eqref{Langevin equation in SR} plays a prominent role and one may expect the axion to quickly travel from one local minimum of its potential to the next one, and thus erase its initial distribution function. The conditions under which this takes place can be assessed by studying stochastic tunneling~\cite{Noorbala:2018zlv, Miyachi:2023fss} between two neighbour minima. Let $\Nt$ denote the tunneling e-folding time starting from $\theta_\uin$, and $P(\Nt,\theta_\uin)$ its distribution function. For simplicity, we consider the case where $\Nt\ll N_H$, hence $H$ can be approximated as a constant. In this case $P(\Nt,\theta_\uin)$ obeys the adjoint Fokker-Planck equation~\cite{Vennin:2015hra, Pattison:2017mbe}
\bea
\label{eq:FP:adj}
\frac{\partial P}{\partial \Nt} 
= -\frac{v_{\mathrm{ax}}'(\theta_\uin)}{3 H^2}\frac{\partial P}{\partial \theta_\uin}+ \frac{H^2}{8\pi^2f^2}\frac{\partial^2 P}{\partial\theta_\uin^2}
\equiv \mathcal{L}^\dagger_{\mathrm{FP}}P\, ,
\eea
where $\mathcal{L}^\dagger_{\mathrm{FP}}$ is the adjoint Fokker-Planck operator. Although this equation has no analytical solution, it gives rise to an ordinary differential equation for the mean tunneling e-folding time $\langle \Nt\rangle$, namely $\mathcal{L}^\dagger_{\mathrm{FP}}\langle \Nt \rangle(\theta_\uin)=-1$, which can be solved as
\begin{equation}
\label{eq:Nt}
    \left\langle N_{\mathrm{t}}\right\rangle\left(\theta_\uin\right) = 8\pi^2\frac{f^2}{H^2}\int_{\theta_\uin}^{2\pi}\mathrm{d}{\theta}_1\int_0^{{\theta}_1}\mathrm{d}{\theta}_2\exp\left\{\frac{8\pi^2}{3}\frac{f^2 m^2}{H^4}\left[\cos\left({\theta}_2\right)-\cos\left({\theta}_1\right)\right]\right\}\, .
\end{equation}
Here, tunneling from $\theta_\uin$ towards either of the two minima located at $\theta=\pm 2 \pi$ is considered, which sets the upper bound of the outer integral and the lower bound of the inner one by imposing $\left\langle N_{\mathrm{t}}\right\rangle\left(\pm 2\pi\right)=0$. When $H^2\ll mf $ and $|\theta_\uin|<\pi$, the argument of the exponential function is large, and is maximal when $\theta_1 = \pi$ and $\theta_2 = 0$. The saddle-point approximation around that point then gives $\langle N_{\mathrm{t}}\rangle = (3/2)\pi H^2/m^2 \ee^{16\pi^2 f^2m^2/(3H^4)}$, taking into account the fact that the $\theta_2$ integral is only over the positive region. This is exponentially larger than the duration of inflation itself if $m<H_\uend$ (if not, the axion becomes massive before the end of inflation, and tunneling is even less effective). This implies that the axion remains trapped around the same local minimum.

In the opposite regime, when $H^2\gg mf $, the argument of the exponential function is negligible, the integrand of \eqref{eq:Nt} is of $\mathcal{O}(1)$ in the whole integration domain and one thus finds that $\langle N_{\mathrm{t}}\rangle$ is of order $f^2/H^2$.\footnote{This coincides with the equilibration e-folding time scale $N_\ueq$ found in \Sec{sec:BeyondEquilibrium} in that case, which is expected.} Recall that this calculation is valid when  $\langle N_{\mathrm{t}}\rangle \ll N_H$, which implies that $H\gg\Hc$ where
\bea
\Hc \equiv \sqrt{f \Hend}\, .
\eea 
Conversely, one can check that $H\gg\Hc$ guarantees that  $H^2\gg mf $ as long as the axion remains light throughout inflation. 

We thus refer to the phase where $H\gg\Hc$ as the ``diffusion-dominated'' regime: during this epoch, tunneling is very efficient (multiple tunneling events occur by the time $H$ changes substantially) and quantum diffusion flattens the distribution of the axion. In this regime indeed, $\Neq \sim f^2/H^2 < N_H$, hence the equilibrium distribution is adiabatically tracked. Since the equilibrium distribution is almost flat under those same conditions, one concludes that if inflation starts at $\Hini>\Hc$, the initial distribution of the axion is quickly erased and is flat at least until $H$ drops below $\Hc$.

In passing, we stress that this conclusion does not rely on starting inflation in the eternal regime. In quadratic inflation indeed, eternal inflation ($\calP_\zeta \gtrsim 1$) takes place when $H$ is above $H_\mathrm{eternal}\sim\sqrt{\Mp H_\uend}$, which is larger than $H_\uc$ as long as $f$ is sub-Planckian.

\subsubsection{Frozen regime}
\label{sec:frozen}

Let us now study the stochastic dynamics of the axion when $H<\Hc$. A first remark is that, in this regime, quantum diffusion plays a negligible role. Indeed, let us assess the field displacement induced by diffusion only, by considering the Langevin equation~\eqref{Langevin equation in SR} in the absence of a drift term. This gives rise to a Gaussian distribution for $\theta$ (or, in the presence of periodic boundary conditions, to an elliptic distribution, see footnote~\ref{PGauss:periodic:elliptic}), where the standard deviation satisfies $\dd\sigma/\dd N=H^2/(8\pi^2f^2\sigma)$ as shown above \Eq{eq:thetacl:sigma:Gauss:quadInfl}. In quadratic inflation, starting from $\sigma=0$ when $H=H_\uin$ this can be solved as
\bea
\label{eq:sigma:quad}
\sigma  =\frac{\sqrt{H_\uin^4-H^4}}{4\pi \Hc^2}\, ,
\eea
which is therefore much smaller than one when $H_\uin\ll H_\uc$. This confirms that quantum diffusion cannot substantially affect the axion distribution below $\Hc$.

Discarding the stochastic noise, the Langevin equation~\eqref{Langevin equation in SR} becomes deterministic and reduces to $\dd\theta/\dd N=-m^2\sin(\theta)/(3H^2)$. Starting from $\theta_\uin$ at $N=N_\uin$, this can be solved as
\bea
\label{eq:classical:SR:anal}
\tan\left(\frac{\theta}{2}\right)=\tan\left(\frac{\theta_\uin}{2}\right)
\exp\left[-\frac{1}{3}\int_{N_\uin}^{N}\frac{m^2(\tilde{N})}{H^2(\tilde{N})}\dd \tilde{N}\right] .
\eea 
When $H>2\pi T_{\mathrm{QCD}}$, from \Eqs{mass temperature}, \eqref{Hubble time} and $\dd N/\dd H=-H/H_{\mathrm{end}}^2$, one has
\bea
\int_{N_\uin}^{N}\frac{m^2(\tilde{N})}{H^2(\tilde{N})}\dd \tilde{N} = \frac{\Lambda^4}{2nf^2 H_\uend^2}\left[\left(\frac{2\pi T_{\mathrm{QCD}
}}{H}\right)^{2n}-\left(\frac{2\pi T_{\mathrm{QCD}
}}{H_\uin}\right)^{2n}\right] .
\eea
Hereafter, $m$ (without time or temperature argument) denotes the mass of the axion at low temperature, \ie $m=\Lambda^2/f$. The above implies that the change in $\ln\tan(\theta/2)$ is suppressed by $m^2/H_\uend^2$, hence it is negligible for axions that remain light until the end of inflation. We call this era, between $H=\Hc$ and $H=2\pi T_{\mathrm{QCD}}$ (or equivalently $H=\Lambda$ since $\Lambda$ and $T_{\mathrm{QCD}}$ are of the same order) the ``frozen regime''. Note that, if the axion is not light at the end of inflation, $m\gtrsim \Hend$, then $\Hc/\Lambda = \sqrt{\Hend/m}\lesssim 1$ and the frozen regime does not exist, which is consistent. 

\subsubsection{Drift-dominated regime}

When $H \leq\Hini <2\pi T_{\mathrm{QCD}} (\sim\Lambda)$, one finds
\bea
\label{eq:integral:drift:dominated}
\int_{N_\uin}^{N}\frac{m^2(\tilde{N})}{H^2(\tilde{N})}\dd \tilde{N} = \frac{\Lambda^4}{f^2 H_\uend^2} \ln\left(\frac{H_\uin}{H}\right)\, .
\eea 
The phase $\Hend<H<\Lambda$ is referred to as the ``drift-dominated regime''.
During this epoch, the evolution of the axion is still suppressed by $m^2/H_\uend^2$, although it receives a logarithmic contribution $\ln(\Lambda/\Hend)$ over the full phase.

\subsubsection{From diffusion to drift}

\begin{figure}[h]
\begin{center}
\begin{tikzpicture}[scale=1]
\draw[white,thick] (0,2) -- (13,2);
\draw[thick,->] (0,0) -- (13,0) node[anchor=north west] {$H$};
\draw[thick] (1,-0.2) -- (1,0.2);
\draw[thick] (3,-0.2) -- (3,0.2);
\draw[thick] (5.5,-0.2) -- (5.5,0.2);
\draw[thick] (8,-0.2) -- (8,0.2);
\draw[thick] (10,-0.2) -- (10,0.2);
\draw[thick] (12,-0.2) -- (12,0.2);
\node[] at (1,-0.5) {$m$};
\node[] at (3,-0.5) {$\Hend$};
\node[] at (5.5,-0.5) {$\;\;\;\;\;\Lambda\sim T_{\mathrm{QCD}}$};
\node[] at (8,-0.5) {$\Hc  = \sqrt{f\Hend}$};
\node[] at (10,-0.5) {$\Hini$};
\node[] at (12,-0.5) {$f$};
\draw[text width=2cm,align=center,color=green,decorate,decoration={brace,raise=0.2cm}]
(10,0) -- (13,0) node[above=0.3cm,pos=0.5] {Pre-inflationary regime};
\draw[text width=2cm,align=center,color=blue,decorate,decoration={brace,raise=0.2cm}]
(8,0) -- (10,0) node[above=0.3cm,pos=0.5] {Diffusion regime};
\draw[text width=2cm,align=center,color=purple,decorate,decoration={brace,raise=0.2cm}]
(5.5,0) -- (8,0) node[above=0.3cm,pos=0.5] {Frozen \\ regime};
\draw[text width=2cm,align=center,color=red,decorate,decoration={brace,raise=0.2cm}]
(3,0) -- (5.5,0) node[above=0.3cm,pos=0.5] {Drift regime};
\draw[text width=2cm,align=center,color=green,decorate,decoration={brace,raise=0.2cm}]
(0,0) -- (3,0) node[above=0.3cm,pos=0.5] {Post-inflationary regime};
\end{tikzpicture}
\caption{Relevant scales in the dynamics of the axion. Inflation starts at $ H = \Hini $ and ends at $ H = \Hend $, where we consider the situation where $m<\Hend$ and $\Hc<\Hini$. Since $H$ decreases during inflation, times flows from right to left. We have placed $\Hini$ below $f$ for illustration only. If $\Hini$ is larger than $f$, PQ breaking occurs during the ``diffusion regime''.}  
\label{fig:energy scale}
\end{center}
\end{figure}
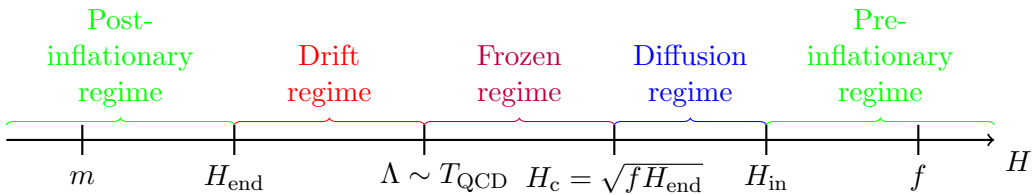

These three different regimes are displayed in the sketch of \Fig{fig:energy scale}. To summarise, above $\Hc$ the axion quickly acquires a flat distribution during the diffusion-dominated phase, between $H=\Hc$ and $H=\Lambda$ that distribution remains frozen, and it may only evolve during the drift-dominated phase at $\Hend<H<\Lambda$.

Inserting \Eq{eq:integral:drift:dominated} with $\Hini=\Lambda$ into \Eq{eq:classical:SR:anal}, the value of the axion at the end of inflation, $\theta_\uend$, is related to its value at the onset of the drift-dominated phase, $\theta_\Lambda$, via
\bea
\tan\left(\frac{\theta_\uend}{2}\right) = \tan\left(\frac{\theta_\Lambda}{2}\right) \left(\frac{\Hend}{\Lambda}\right)^{\frac{\Lambda^4}{3f^2\Hend^2}}\, .
\eea 
This defines the function $\theta_\uend(\theta_\Lambda)$, from which the distribution function of the axion at the end of inflation is given by
\bea
\label{eq:P:thetaend}
P\left(\theta_\uend\right) =  P\left(\theta_\Lambda\right)\left\vert \frac{\dd\theta_\Lambda}{\dd\theta_\uend}\right\vert
= \frac{1}{2\pi} \frac{1}{X+\left(\frac{1}{X}-X\right)\sin^2\left(\frac{\theta_\uend}{2}\right)}\, ,
\eea 
where we have used that $P(\theta_\Lambda)$ is flat and we have introduced $X=\left(\frac{\Hend}{\Lambda}\right)^{\frac{\Lambda^4}{3f^2\Hend^2}}$. This distribution is, obviously, quite different from the instantaneous equilibrium~\eqref{stationary solution Fokker-Planck equation} at the end of inflation. When $\Hc\gg\Lambda \ln^{1/4}(\Lambda/\Hend)$, $X$ is close to one and $P(\theta_\uend) $ is almost flat. In the opposite regime, $X\ll 1$ and the distribution is peaked around the minimum of the potential at $\theta=0$. 

So far we have assumed that $ \Hend $ is larger than the axion mass at zero temperature, and that inflation starts in the diffusion regime where initial conditions are quickly erased by quantum diffusion, $\Hini>\Hc$. Below, we discuss the possibility that these assumptions may be broken.

\subsection{Beyond slow roll}
\label{sec:BeyondSlowRoll}

As inflation proceeds, $H$ decreases. If it becomes of order the mass of the axion before the end of inflation, the slow-roll approximation for the axion dynamics breaks down, and the above considerations have to be revised. 

We first note that, when $H\sim \Lambda$, $m/H\sim \Lambda/f$, which is very small (recall that $\Lambda \sim 100\, \MeV$ and below we will see that $f$ has to be above $10^{10}\, \GeV$ for the axion to comprise all of dark matter). Therefore, the axion is always light in the diffusion and frozen regimes, that is when $H>\Lambda$, and the conclusion drawn in \Sec{sec:diffusion:dominated} that its distribution is flat when exiting that phase remains valid. Moreover, since the amplitude of the stochastic noise decreases as the mass of the axion increases relative to the Hubble expansion rate, the conclusion that quantum diffusion plays a negligible role when $H<\Lambda$ remains valid too.

Possible violations of slow roll may thus only affect the epoch dominated by classical drift, $H<\Lambda$. During this phase the mass of the axion is constant and the axion follows the Klein-Gordon equation
\bea
\label{eq:KG}
\frac{\dd^2\theta}{\dd N^2}
+\left[3-\epsilon_1(N)\right]\frac{\dd\theta}{\dd N}+\frac{m^2}{H^2(N)}\sin(\theta)=0\, ,
\eea
where we recall that $\epsilon_1=H_\uend^2/H^2$ in quadratic inflation. By solving this equation numerically, one can reconstruct the function $\theta_\uend(\theta_\Lambda)$ and evaluate $P(\theta_\uend) = P(\theta_\Lambda) \vert \dd \theta_\Lambda/\dd\theta_\uend\vert$.\footnote{Instead of computing the Jacobian $\dd \theta_\Lambda/\dd\theta_\uend$ by differentiating the numerical solution of \Eq{eq:KG}, it is numerically more efficient to evaluate it as $\delta\theta_\Lambda/\delta\theta_\uend$, where the axion infinitesimal displacement  $\delta\theta$ is obtained from solving the linearised Klein-Gordon equation
\begin{equation}
\label{Klein-Gordon delta theta}
    \frac{\dd^2\delta\theta}{\dd N^2} + \left[3-\epsilon_1(N)\right]\frac{\dd\delta\theta}{\dd N} + \frac{m^2}{H(N)^2}\cos\left(\theta\right)\delta\theta = 0
\end{equation}
where $\theta$ follows \Eq{eq:KG}.} 
\begin{figure}
\centering
\includegraphics[width=0.49\textwidth]{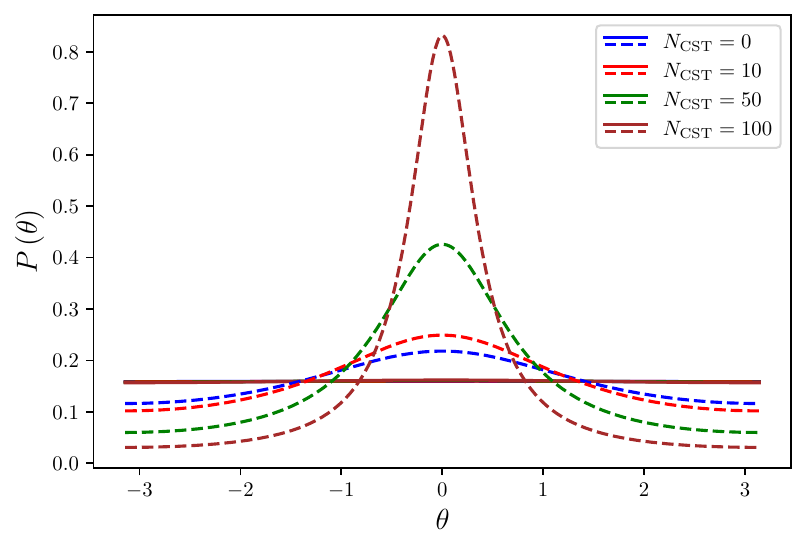}
\includegraphics[width=0.49\textwidth]{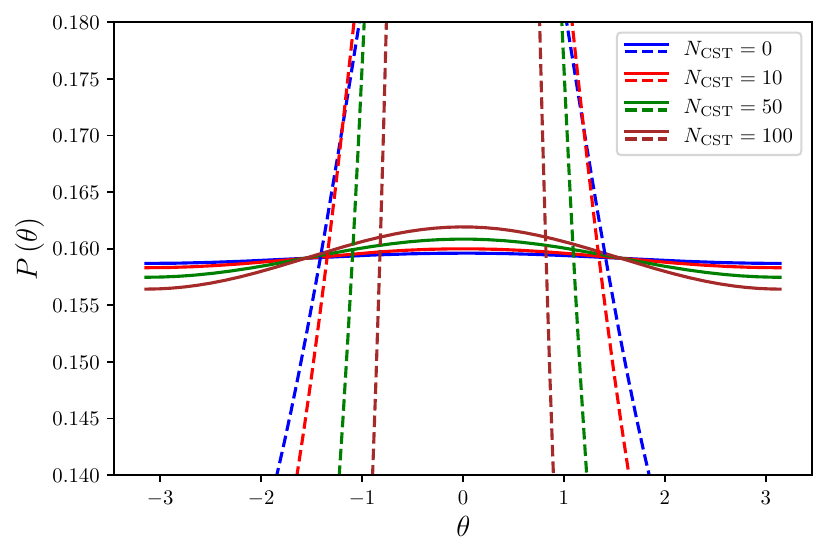}
			\caption{Distribution of the misalignment angle at the end of inflation for $ \Hend = 50 m $ (solid curves) and $ \Hend = 5 m $ (dashed curves). An additional phase of plateau inflation is added before inflation ends, the duration of which, $\NCST$, is labeled with different colours. We assume that inflation starts at sufficiently high energy, $\Hini>\Hc$, so that quantum diffusion flattens the initial distribution. The right panel shows the same plots as the left but with a narrower range of the vertical axis.} 
			\label{fig:distribution function NCST}
\end{figure}
In practice, we make use of \Eqs{eq:classical:SR:anal} and~\eqref{eq:integral:drift:dominated} as long as $H>10m$,  and \Eq{eq:KG} afterwards (we have checked that the same results are recovered if one uses \Eq{eq:KG} all along). The distribution one obtains is displayed in \Fig{fig:distribution function NCST} for $\Hend=50 m$ (blue solid curve) and for $\Hend=5m$ (blue dashed curve). In the latter case, the slow-roll approximation breaks down towards the end of inflation.

Finally, we note that the validity of the formula $P(\theta_\uend) = P(\theta_\Lambda) \vert \dd \theta_\Lambda/\dd\theta_\uend\vert$ requires that there is a one-to-one relationship between $\theta_\Lambda$ and $\theta_\uend$. Otherwise, a sum over all values of $\theta_\Lambda$ leading to a given $\theta_\uend$ has to be performed. This is the case if the axion oscillates at the bottom of its potential before inflation ends, since then the function $\theta_\uend(\theta_\Lambda)$ is not monotonic. In \App{app:Oscillations}, we show that oscillations take place if and only if $m>3\Hend$. In what follows, we will check that the axion can never provide the observed amount of dark matter in this regime, so in practice oscillations do not need to be accounted for.

\subsection{Dependence on the initial conditions}

Above we have shown that if inflation starts at $\Hini\gg\Hc$, then initial conditions are erased and the axion swiftly acquires a flat distribution. To assess the role of initial conditions when $\Hini\lesssim\Hc$, in this section we consider the situation where the axion has an initial Dirac distribution, $P_\uin(\theta)=\delta(\theta-\theta_\uin)$, since this ``maximally'' contrasts with a flat profile. 

If $\Hini>\Hc$, a first phase takes place where the dynamics of the axion is dominated by quantum diffusion, until $H=\Hc$. In this regime, \Eq{eq:PGauss:periodic:elliptic} applies, where $\sigma$ is given by \Eq{eq:sigma:quad}. 
One recovers that, if $\Hini\gg \Hc$, $\sigma^2\gg 1$ and a quasi-flat distribution is quickly reached. Otherwise, the distribution of the axion may still be peaked when it exits the diffusion-dominated regime. 

\begin{figure}
\centering
\includegraphics[width=0.49\textwidth]{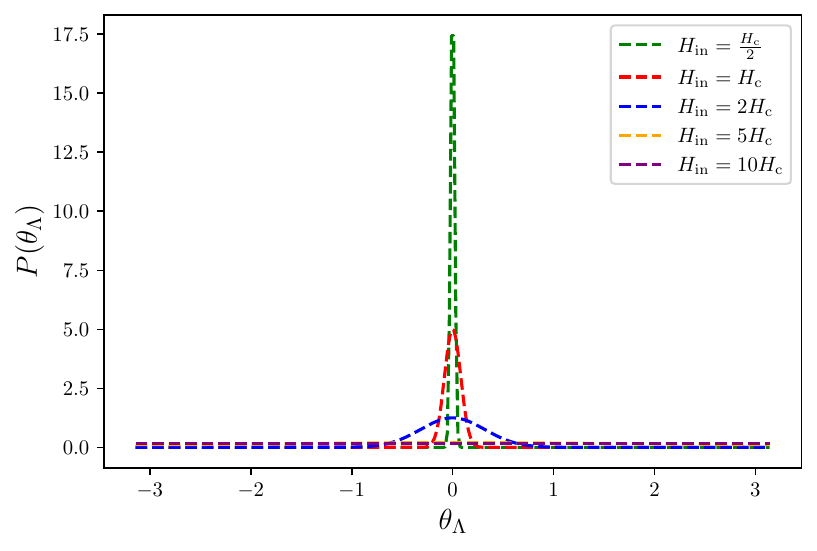}
\includegraphics[width=0.49\textwidth]{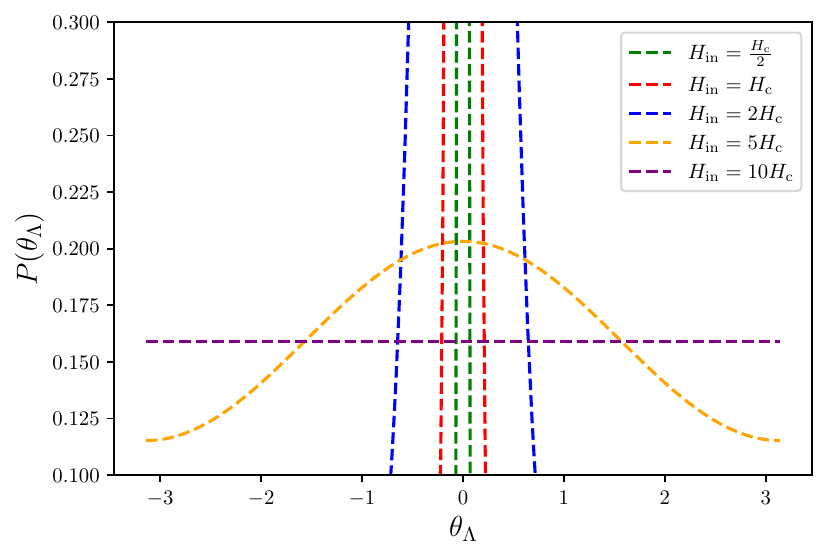}
			\caption{Distribution of the misalignment angle $\theta_\Lambda$ when $ H $ crosses $ \Lambda = 0.1$~$ \GeV $, starting from a Dirac distribution centred at $ \theta = 0 $ for different values of $ \Hini $ and for $ \Hc  = 10 \, \GeV $. As long as $ \Hend \gg m $, this is a good proxy for the distribution at the end of inflation. The right panel zooms in on the left one to highlight the small discrepancy between the yellow and the purple curves.} 
			\label{fig:theta 3 solution}
\end{figure}

Its evolution is then driven by the classical drift, and it can be processed using \Eq{eq:P:thetaend}, or the approach detailed in \Sec{sec:BeyondSlowRoll} if slow roll breaks down, where in both cases $P(\theta_\Lambda)$ needs to be replaced with \Eq{eq:PGauss:periodic:elliptic} rather than assumed to be flat. 
The result is displayed in \Fig{fig:theta 3 solution} for a few values of $\Hini$. One can check that, when $\Hini\gtrsim 10 \Hc$, the distribution of the axion is flat, while it is more narrow for smaller values of $\Hini$.

Therefore, the final distribution is independent of the initial one if $\Hini>\Hc$, which corresponds to inflation lasting more than $N=\Hc^2/\Hend^2=f/\Hend$ \efolds. If $\Hc<\Hini$, predictions depend on the initial state of the axion. However, even in that case, if the initial distribution of the axion is flat, our conclusions remain unchanged. Flat initial distributions may be natural given the shift symmetry of the axion prior to the symmetry breaking.

\subsection{Adding a phase of plateau inflation}
\label{sec:plateau}

As argued in \Sec{sec:BeyondEquilibrium}, the bulk of the inflationary era is likely to be described by large-field potentials of the type~\eqref{eq:quad:pot:phi}, rather than by plateau potentials. However, CMB measurements~\cite{Planck:2018jri} imply that during $50-60$ \efolds before the end of inflation, the inflationary potential is of the plateau type, and in this section we estimate how the above results change when one considers this last stage of plateau inflation. 

We suppose that plateau inflation occurs at $H_{\mathrm{p}}\simeq\Hend<\Hc$ hence \Eq{eq:classical:SR:anal} applies, and reduces to
\bea 
\tan\left(\frac{\theta_\uend}{2}\right) = \tan\left(\frac{\theta_{\mathrm{p}}}{2}\right)\ee^{-\frac{m^2}{3H_\uend^2} \NCST}
\eea 
when $H$ is quasi constant. Here, $\theta_{\mathrm{p}}$ denotes the value of the axion at the onset of the plateau phase, which lasts for $\NCST$ \efolds. One can see that, as long as the axion is sufficiently light, $m\ll \Hend/\sqrt{\NCST}$, the plateau phase has negligible impact on its distribution at the end of inflation. In the opposite regime, $m\gg \Hend/\sqrt{\NCST}$, one has $N_\ueq\ll \NCST$, see \Eq{eq:Neq}, hence the axion reaches the equilibrium distribution~\eqref{stationary solution Fokker-Planck equation}. 
The distribution $P(\theta_\uend)$ is displayed in \Fig{fig:distribution function NCST} for a few values of $\NCST$. One can check that, when $\NCST$ is of order $100$, the final phase of plateau inflation has negligible impact as soon as $\Hend\gtrsim 50 m$. As we will see below, for the axion to account for dark matter at these scales, one must have $f\sim 10^{12}\, \GeV$ and $\Hend<10^8\, \GeV$, hence it would then take more than $10^{11}$ \efolds of plateau inflation to reach the late-time equilibrium, a very large value indeed. 

\subsection{Alternative large-field completions}
\label{sec:alternative:completions}
 
Above we have assumed that inflation proceeds in a quadratic potential prior to the plateau phase, arguing that mass terms generically arise as radiative corrections for quantum fields in expanding backgrounds. However, other terms may be generated with higher powers, for instance quartic terms, that would dominate the inflaton's potential at large-field values. In this section we thus examine the possibility of other large-field completions, and consider the generic potential 
\bea
V(\phi)\propto \phi^p\, ,
\eea
for which~\cite{Martin:2013tda} $H(N) = \Hend [1+4\left(\Nend-N\right)/p]^{p/4}$. For $p=2$, one can check that this reduces to \Eq{Hubble time}. The e-folding time scale describing the background expansion reads $N_H=1/\epsilon_1=(H/\Hend)^{4/p}$. When $H\gg \sqrt{mf}$, one still has $N_\ueq\sim f^2/H^2$, hence the condition $N_\ueq<N_H$ to be in the diffusion-dominated regime, see \Sec{sec:diffusion:dominated}, reads $H>H_\uc$ with
 \bea
 \label{eq:Hc:gen}
 H_\uc = \Hend \left(\frac{f}{\Hend}\right)^{\frac{1}{1+\frac{2}{p}}}\, .
 \eea
 Therefore, if inflation starts at $\Hini>\Hc$, the distribution of the axion quickly becomes flat. When $\Hini<\Hc$, the importance of quantum diffusion can be assessed by considering the standard deviation acquired by the axion under the influence of stochastic noise only, and one finds
 \bea
 \sigma^2 = \frac{1}{8\pi^2\left(1+\frac{2}{p}\right)}\left[\left(\frac{H_\uin}{\Hc}\right)^{2+\frac{4}{p}}-\left(\frac{H}{\Hc}\right)^{2+\frac{4}{p}}\right] ,
 \eea
which generalises \Eq{eq:sigma:quad}. From this expression, one can check that quantum diffusion plays a negligible role when $H\ll \Hc$, so this regime can be treated by considering classical drift only. The same arguments as those presented in \Sec{sec:frozen} can thus be made. In particular, \Eq{eq:classical:SR:anal} still applies, where now
 \bea
 \int_{N_\uin}^{N}\frac{m^2(\tilde{N})}{H^2(\tilde{N})}\dd \tilde{N} =
 \begin{cases}
\frac{1}{2(n+1-\frac{2}{p})}\left( \frac{\Lambda}{\Hc} \right)^4 \left(\frac{2\pi T_{\mathrm{QCD}}}{\Hc}\right)^{\frac{4}{p}-2} 
 \\ \quad \times 
\left[\left(\frac{2\pi T_{\mathrm{QCD}}}{H}\right)^{2n+2-\frac{4}{p}}-\left(\frac{2\pi T_{\mathrm{QCD}}}{H_\uin}\right)^{2n+2-\frac{4}{p}}\right] 
\quad &\text{if}\quad H_\uin \geq H>2\pi T_{\mathrm{QCD}}\, ,
\\
\frac{\Lambda^4}{\left(2-\frac{4}{p}\right) f^2 \Hend^2} \left[\left(\frac{H_\uend}{H}\right)^{2-\frac{4}{p}} - \left(\frac{\Hend}{H_\uin}\right)^{2-\frac{4}{p}}\right]
\quad &\text{if}\quad H \leq H_\uin \leq 2\pi T_{\mathrm{QCD}}\, .
\end{cases}
 \eea
These expressions reveal that $p=2$ is a singular case if $H \leq 2\pi T_{\mathrm{QCD}}$. For $p>2$, the above integral remains negligible when $H>T_{\mathrm{QCD}}\sim \Lambda$, hence the regime $\Lambda<H<\Hc$ is frozen as in the quadratic case. Between $H=\Lambda$ and $H=\Hend$, the above integral is of order $m^2/\Hend^2$, hence the distribution for the axion remains flat unless the axion stops being light before the end of inflation, and one recovers exactly the same conclusion as in quadratic inflation.

For $p<2$, two cases need to be distinguished. If $p>2/(n+1)$, the above integral with $H=\Lambda$ is of order $(\Lambda/\Hc)^{4/p+2}$ when $H_\uin>\Lambda$, hence it remains small and this regime is still frozen. If $p<2/(n+1)$, the integral rather scales as $(\Lambda/\Hc)^{4/p+2}(H_\uin/\Lambda)^{4/p-2n-2}$. If one evaluates it from the end of the diffusion-dominated phase, $H_\uin=\Hc$, it is thus of order $(\Hc/\Lambda)^{-2n-4}$, hence it is also small. Thus the distribution of the axion does not narrow down in the phase $\Lambda<H<\Hc$ in both cases. Between $H=\Lambda$ and $H=\Hend$, the integral scales as $(\Lambda/\Hc)^{2+4/p}$, which is small as long as the axion remains light until the end of inflation, so one reaches similar conclusions as in quadratic inflation in those last phases.

To summarise, the results derived above in quadratic inflation qualitatively generalise to any large-field potential $V(\phi)\propto \phi^p$, where $\Hc$ is now given by \Eq{eq:Hc:gen}. In particular, the distribution of the axion is found to be flat at the end of inflation unless the axion stops being light before then.\\

Another possibility worth discussing is that inflation proceeds in hilltop potentials. Although quadratic-hilltop inflation is now strongly disfavoured by CMB measurements~\cite{Martin:2013nzq}, quartic-hilltop inflation, 
\bea
V(\phi) \propto 1-\left(\frac{\phi}{\mu}\right)^4\, ,
\eea
is still compatible with the data for $\mu\gg\Mp$. In those models, $H$ can be approximated as constant at early time, so the axion does reach the equilibrium distribution provided inflation lasts long enough. However, similarly to the case of plateau inflation discussed at the end of \Sec{sec:BeyondEquilibrium}, this requires to start inflation in the eternal regime. Indeed, in quartic-hilltop inflation one has~\cite{Martin:2013tda} $\epsilon_1\propto(N_\uend-N)^{-3}$ when $N_\uend-N\gg 1$. The amplitude of the power spectrum of curvature perturbations thus satisfies
\bea
\calP_{\zeta} \simeq \calP_{\zeta}(k_0) \left(\frac{ N_\uend-N}{N_\uend-  N_*} \right)^3\, ,
\eea
where $\calP_{\zeta}(k_0)\simeq 2.2\times 10^{-9}$ is  the amplitude of the spectrum measured at the CMB pivot scale $k_0=0.05\,\Mpc^{-1}$, for which $N_\uend-N_*\sim 50$. As a consequence, for inflation to start away from the eternal regime, $\calP_{\zeta,\uin}\ll 1$, one requires $N_\uend - N_\uin\ll 4\times 10^4$. This duration must be larger than $N_\ueq$ for the equilibrium to be reached, which implies that $H\ll 200 m$ if $H\ll\Lambda$ (for which $N_\ueq = H^2/m^2$) and $f\ll 15 H$ if $H\gg\Lambda$ (for which $N_\ueq = 16\pi^4 f^2/H^2$). Below we show that these conditions exclude the possibility for the axion to constitute all of dark matter (see \Fig{fig:EB-2d-1plot}).

\section{Stochastic window for the axion}
\label{sec:Stochastic:Window}

Having explained how the distribution of the axion can be computed at the end of inflation, we now study its phenomenological consequences, and identify the regions of parameter space where the axion may constitute dark matter while passing isocurvature constraints. 

\subsection{Axion dark matter}

If the fractional energy contained in the axion today is identified with the dark-matter abundance, it is given by~\cite{Ballesteros:2016xej,Sikivie:2006ni,Bae:2008ue}
\bea
\label{eq:h2OmegaDM}
h^2 \Omega_{\mathrm{DM}} = b \theta_\uend^2 \left[1-\ln\left(1-\frac{\theta_\uend^2}{\pi^2}\right)\right]^\beta \left(\frac{f}{\hat{f}}\right)^\beta
\eea 
where $b=0.195$, $\hat{f}=10^{12}\, \GeV$, $\beta=1.184$ and $h=H_0/(100\, \mathrm{km}\, \mathrm{s}^{-1}\, \mathrm{Mpc}^{-1})$ is the rescaled Hubble parameter at present and we recall that $\theta_\uend$ denotes the misalignment angle at the end of inflation. When $f$ increases, the axion remains light, hence frozen, for a longer period of time, hence its abundance today is larger, which explains why $\Omega_{\mathrm{DM}}$ increases with $f$. Likewise, if $\theta_\uend$ increases, more energy is stored in the axion at the end of inflation, which also explains why $\Omega_{\mathrm{DM}}$ increases with $\theta_\uend$. As a consequence, in order to account for the observed value of $\Omega_{\mathrm{DM}}$, $\theta_\uend$ has to be larger as $f$ decreases.

However, $f$ cannot be arbitrarily small, otherwise the axion yields unacceptably large isocurvature perturbations. The amplitude of the isocurvature power spectrum, $\calP_S$, can be estimated as~\cite{Kobayashi:2013nva}
\bea
\sqrt{\calP_S(k)} = \frac{\delta\rho_\theta}{\rho_\theta} = \frac{\partial\ln\Omega_{\mathrm{DM}}}{\partial\theta_*}\delta\theta_*\, ,
\eea 
where $\rho_\theta$ is the energy density contained in the axion today and $\delta\rho_\theta$ is its fluctuation. In the second expression, we have assumed that the axion constitutes all of dark matter, hence $\rho_\theta=\rho_{\mathrm{DM}}$, and $\theta_*$ corresponds to the misalignment angle when $k$ crosses out the Hubble radius during inflation. Below we will see that isocurvature constraints are relevant only when the axion remains extremely light until the end of inflation, hence one may replace $\theta_*$ by $\theta_\uend$ in the above expression, and  $\delta\theta_*=H_*/(2\pi f)\simeq \Hend/(2\pi f)$. One thus finds
\bea
\label{eq:calPS}
\sqrt{\calP_S(k)} \simeq \frac{\partial\ln\Omega_{\mathrm{DM}}}{\partial\theta_\uend} \frac{\Hend}{2\pi f}\, .
\eea 
The amplitude of isocurvature perturbations is usually measured through the ratio
\bea 
\label{eq:alpha}
\alpha = \frac{\calP_S(k_0)}{\calP_\zeta(k_0)}\, ,
\eea 
where, as already mentioned, $\calP_\zeta(k_0)\simeq 2.2\times 10^{-9}$ is the power spectrum of curvature perturbations. The current constraint~\cite{Planck:2018jri} $\alpha<0.041$ thus leads to an upper bound on $\calP_S(k_0)$, \ie on the right-hand side of \Eq{eq:calPS}.

For a given value of $f$, the misalignment angle $\theta_\uend$ that produces the observed abundance of dark matter, $h^2\Omega_{\mathrm{DM}}\simeq 0.12$, can be obtained from \Eq{eq:h2OmegaDM}. This value for $\theta_\uend$ may be more or less likely, depending on the probability distribution $P(\theta_\uend)$ according to which it is drawn, and which was computed in \Sec{sec:distribution end inflation}. This, in turn, places probabilistic constraints on the parameters of that distribution, $\Hend$ and $f$, which can be derived using the following Bayesian framework. 

\subsection{Bayesian inference}

In this work we employ Bayesian inference methods to reconstruct the constraints on $f$ and $\Hend$ induced by measurements of the dark-matter abundance and by upper bounds on the amplitude of isocurvature perturbations.

Given a model $\mathcal{M}_\alpha$, described by a set of parameters $p_{\alpha i}$, the probability to observe a certain data $D$ is called the likelihood function and is denoted $P(D\vert \mathcal{M}_\alpha,p_{\alpha i})$. Using Bayes' theorem, it can be related to the probability that, having observed the data $D$, the underlying values of the model's parameters are $p_{\alpha i}$,
\bea
P\left(p_{\alpha i}\vert D,\mathcal{M}_\alpha\right) = \frac{P\left(D\vert \mathcal{M}_\alpha,p_{\alpha i}\right)P\left(p_{\alpha i}\vert\mathcal{M}_\alpha\right)}{P\left(D\vert \mathcal{M}_\alpha\right)}\, ,
\eea 
which is called the posterior distribution on the parameters $p_{\alpha i}$. In this expression, $P(p_{\alpha i}\vert\mathcal{M}_\alpha)$ is the prior distribution on the parameters $p_{\alpha i}$, \ie prior to performing measurements, and $P(D\vert \mathcal{M}_\alpha)$ is the so-called Bayesian evidence of the model $\mathcal{M}_\alpha$. By the basic property of conditional probabilities one has
\bea
P\left(D\vert\mathcal{M}_\alpha\right) = \int \dd p_{\alpha i} P\left(D\vert \mathcal{M}_\alpha,p_{\alpha i}\right)P\left(p_{\alpha i}\vert\mathcal{M}_\alpha\right)\, ,
\eea
and this guarantees that the posterior distribution is properly normalised. Using Bayes' theorem again, the posterior probability of the model $\mathcal{M}_\alpha$ itself is given by
\bea
\label{eq:posterior}
P\left(\mathcal{M}_\alpha \vert D\right ) = \frac{P\left(D\vert\mathcal{M}_\alpha\right) P\left(\mathcal{M}_\alpha\right)}{P(D)}\, ,
\eea 
where $P(\mathcal{M}_\alpha)$ is the prior belief in model $\mathcal{M}_\alpha$, and $P(D)=\sum_\alpha P(D|\mathcal{M}_\alpha)P(\mathcal{M}_\alpha)$ is a normalisation factor. For a review on the Bayesian approach to model comparison, see \eg \Refa{Trotta:2008qt}.

The above formalism can be applied to the setup discussed in this work as follows. A ``model'' corresponds to a tuple of values for $\Hend$ and $f$. Its prior, $P(\Hend,f)$ is thus nothing but the prior distribution of these two quantities. Each model has one parameter, $\theta_\uend$, and the object of \Sec{sec:distribution end inflation} was precisely to compute its prior distribution, $P(\theta_\uend \vert \Hend,f)$. The likelihood function, $P(D \vert \Hend, f, \theta_\uend)$, implements the observational constraints on the abundance of dark matter $h^2\Omega_{\mathrm{DM}}$ and the amount of isocurvature perturbations $\alpha$. In practice, we assume these two cosmological parameters to be uncorrelated, 
\bea
\label{eq:likelihood}
P\left(D \vert \Hend,f,\theta_\uend\right) = \Theta\left[{\alpha}_{\mathrm{m}} - \alpha\left(\Hend,f,\theta_\uend\right)\right]
\frac{\ee^{-\frac{\left[h^2\Omega_{\mathrm{DM}}\left(\Hend,f,\theta_\uend\right)-h^2\bar{\Omega}_{\mathrm{DM}}\right]^2}{2 \sigma_{h^2\Omega_{\mathrm{DM}}}}}}{\sqrt{2\pi \sigma_{h^2\Omega_{\mathrm{DM}}}}}\, ,
\eea
and consider $D = ({\alpha}_{\mathrm{m}}, h^2\bar{\Omega}_{\mathrm{DM}}, \sigma_{h^2\Omega_{\mathrm{DM}}})$.
In this expression, for simplicity we model the likelihood for isocurvature perturbations by a Heaviside step function $\Theta$, where ${\alpha}_{\mathrm{m}}=0.041$ is the upper bound mentioned above, and $\alpha(\Hend,f,\theta_\uend)$ is obtained from \Eqs{eq:calPS}-\eqref{eq:alpha}. We also model the measurement of dark-matter abundance, $h^2\Omega_{\mathrm{DM}}=0.12\pm 10^{-3}$~\cite{Planck:2018vyg}, by a Gaussian likelihood with $h^2\bar{\Omega}_{\mathrm{DM}}=0.12$ and $\sigma_{h^2\mathrm{DM}}=10^{-3}$, and $h^2\Omega_{\mathrm{DM}}\left(\Hend,f,\theta_\uend\right)$ is given by \Eq{eq:h2OmegaDM}.

Combining the above results, the posterior distribution \eqref{eq:posterior} for $\Hend$ and $f$ can be written as 
\bea
\label{eq:posterior:Hend_f}
P\left(\Hend,f\vert D\right) = \int_{0}^{2\pi} \dd\theta_\uend P\left(D\vert \Hend,f,\theta_\uend\right) P\left(\theta_\uend \vert \Hend,f\right) \frac{P\left(\Hend,f\right)}{P(D)}\, .
\eea 
In this expression, we recall that the likelihood function $P\left(D\vert \Hend,f,\theta_\uend\right)$ is given in \Eq{eq:likelihood}; the prior distribution on $\theta_\uend$, $P\left(\theta_\uend \vert \Hend,f\right)$, is derived according to the procedure detailed in \Sec{sec:distribution end inflation}; $P(\Hend,f)$ is the prior distribution on $\Hend$ and $f$, which in practice will be taken of the Jeffrey's kind, \ie uniform in the logarithm of the parameters; and $P(D)$ is a normalisation factor. In practice, it is numerically convenient to trade the integral over $\theta_\uend$ in \Eq{eq:posterior:Hend_f} into an integral over $h^2\Omega_{\mathrm{DM}}$, using \Eq{eq:h2OmegaDM} for fixed values of $\Hend$ and $f$. In figures below we plot 
\bea
\label{eq:posterior:logHend_logf}
P\left(\log\Hend,\log f\vert D\right) = \Hend f \, P\left(\Hend,f\vert D\right)\,,
\eea
which satisfies $P\left(\log\Hend,\log f\vert D\right) \dd(\log\Hend) \dd(\log f) = P\left(\Hend,f\vert D\right) \dd\Hend \dd f$.

\subsection{Constraints on the parameters}

\begin{figure}[t]
\begin{center}
\includegraphics[width=1\textwidth]{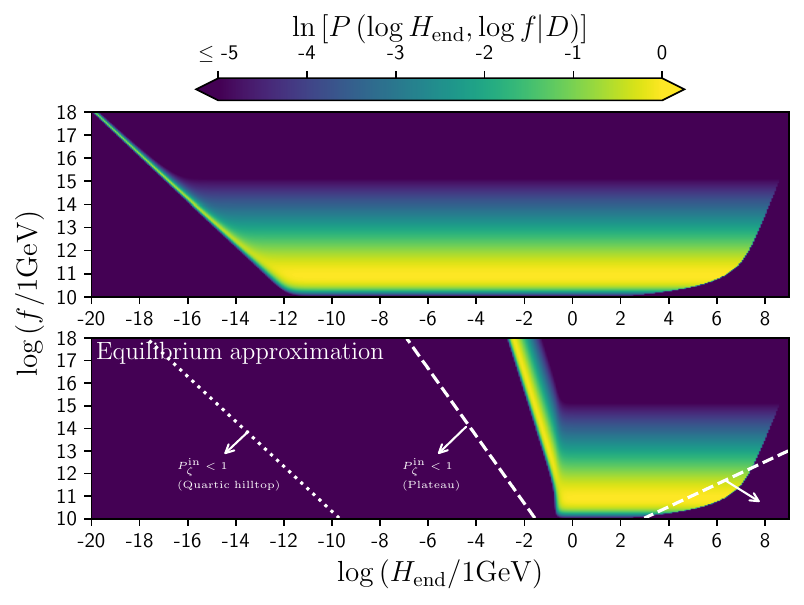}
\caption{Posterior distribution as a function of $\log f $ and $ \log\Hend $, defined in \eqref{eq:posterior:logHend_logf}. The top panel is obtained with $ \NCST = 0 $, while the bottom panel corresponds to assuming that the equilibrium distribution holds at the end of inflation. In this case, the dashed and dotted lines delineate the region where the equilibrium distribution can be reached without starting in the eternal regime, for infinite plateau and quadratic hilltop potentials respectively.}
\label{fig:EB-2d-1plot}
\end{center}
\end{figure}

In \Fig{fig:EB-2d-1plot}, the posterior distribution~\eqref{eq:posterior:logHend_logf} is displayed for two different choices of the distribution $P\left(\theta_\uend \vert \Hend,f\right)$: one using the results of \Sec{sec:distribution end inflation} to compute the probability distribution of the axion at the end of inflation (top panel), and the other assuming stochastic equilibrium~\eqref{stationary solution Fokker-Planck equation} at the end of inflation, as usually done in the literature~\cite{Graham:2018jyp, Takahashi:2018tdu} (bottom panel). As mentioned above, a logarithmically-flat prior is employed for both $\Hend$ and $f$, to ease the interpretation in a logarithmically-scaled figure, and the posterior distribution is normalised to its maximal value. In this way, the colour bar can be interpreted in terms of the Jeffreys' scale, where we recall that ``favoured'' models correspond to $-1\leq \ln P$, ``weakly disfavoured'' models to $-2.5\leq P<-1$, ``moderately disfavoured'' models to $-5\leq P<-2.5$ and ``strongly disfavoured'' to $P<-5$. In this sense, values of $\Hend$ and $f$ falling in the dark blue region can be considered as ruled out. Several remarks are in  order.

\subsubsection*{Joint constraints}

When $\Hend \gg \Lambda=0.1\GeV$ (and $\Hini\gg\Hc\gg\Hend$), from \Sec{sec:distribution end inflation} it is clear that the axion has a flat distribution at the end of inflation. In this regime, the equilibrium profile~\eqref{stationary solution Fokker-Planck equation} is also flat, hence both coincide. Let us stress that, in this case, there is no adiabatic relaxation towards the equilibrium, and the agreement between the two profiles shall be seen as purely coincidental. In practice, it implies that the posterior distribution on $\Hend$ and $f$ is the same for both distributions in this regime. The likelihood does not depend on $\Hend$ as far as the isocurvature constraint is not relevant, given that neither $P(\theta_\uend|\Hend,f)$, which is flat, nor $\Omega_{\mathrm{DM}}$, which is given by \Eq{eq:h2OmegaDM}, depend on $\Hend$. Since $\theta_\uend$ is of order one, from \Eq{eq:h2OmegaDM} $\Omega_{\mathrm{DM}}$ is also of order one when $f$ is of order $\hat{f}=10^{12}\, \GeV$, around which the preferred values of $f$ indeed lie.

Since $\theta_\uend$ is of order one in the regime, $\partial\ln\Omega_{\mathrm{DM}}/\partial\theta_\uend$ in \Eq{eq:calPS} is also of order one, hence the isocurvature constraint roughly translates into $\Hend <10^{-5} f$, which leads to $\Hend<10^7\, \GeV$ when $f\sim \hat{f}$. This explains why values of $\Hend$ above this threshold are forbidden, since they lead to isocurvature perturbations with too large amplitude. In this regime, $m/\Hend=\Lambda^2/(\Hend f)<10^{-21}$, so the axion remains very light until the end of inflation, which validates a posteriori the assumption made above \Eq{eq:calPS}.

When $\Hend \ll \Lambda=0.1\GeV$ the equilibrium profile~\eqref{stationary solution Fokker-Planck equation} is peaked and almost Gaussian, with a variance $\langle \theta_\uend^2\rangle \simeq 3\Hend^4/(8\pi^2\Lambda^4)$. In this limit \Eq{eq:h2OmegaDM} reduces to $h^2\Omega_{\mathrm{DM}} \simeq b \theta_\uend^2 (f/\hat{f})^\beta$, so to reach the observed abundance of dark matter one must have $f\sim 10^{13}\GeV (\Hend/\Lambda)^{-4/\beta}$. This explains the tilted part of the axion window in the bottom panel of \Fig{fig:EB-2d-1plot}. In practice, values of $\Hend$ smaller than $10^{-3}\GeV$ lead to super-Planckian values for $f$ and are thus excluded.

However, as long as the axion remains light throughout inflation, we have seen in \Sec{sec:distribution end inflation} that its actual distribution is approximately flat at the end of inflation, hence the dark-matter constraint leads to $f\sim \hat{f}$ even when $\Hend\ll \Lambda$, as long as $m\ll \Hend$, \ie $ \Hend \gg \Lambda^2/f  $. For the lowest values allowed for $f$, that are of order $10^{10}\GeV$ in \Fig{fig:EB-2d-1plot}, this regime corresponds to $\Hend\gg 10^{-12}\GeV$. It thus extends well below the scale $\Lambda$, where the conclusions drawn from the equilibrium distribution (which is peaked) are invalid. 

When $\Hend\ll 10^{-12}\GeV$, the axion cannot be considered as light at the end of inflation, and its distribution is not flat anymore. This corresponds to the tilted part of the axionic window in the top panel of \Fig{fig:EB-2d-1plot}, where $m/\Hend$ is of order one, \ie $f\sim \Lambda^2/\Hend$. This window extends all the way up to the Planck scale, $f\sim 10^{18}\GeV$, where the corresponding value of $\Hend\sim 10^{-20}\GeV$ coincidentally matches the one at big-bang nucleosynthesis.\footnote{Big-bang nucleosynthesis proceeds at temperatures of order $100\,\MeV$, which coincides with $\Lambda$, and using the Friedmann equation, $H^2=\rho/(3\Mp^2)\sim T^4/(3\Mp^2)$, this corresponds to $H\sim 10^{-20}\GeV$.} Therefore, the requirement that inflation proceeds before big-bang nucleosynthesis, and the requirement that the axion symmetry breaking scale $f$ needs to be sub-Planckian, happen to yield the same lower bound on the energy scale of inflation, although these two conditions are physically disconnected.

In passing, in the bottom panel of \Fig{fig:EB-2d-1plot}, we display the regions of parameter space where the equilibrium distribution can be reached without starting inflation in the eternal regime, for infinite plateau (for $\gamma=1$) and quartic hilltop models, see the discussions at the end of \Secs{sec:BeyondEquilibrium} and \ref{sec:alternative:completions} respectively. For quartic hilltop models, the conclusion is that inflation needs to start in the eternal regime. For infinite plateau models, only a small region of parameter space seems compatible with non-eternal inflation. However, in this region, we have checked that $P_\zeta^{\mathrm{in}}$ is still high, requiring a non-perturbative description even though $P_\zeta^{\mathrm{in}} < 1$. Thus the same conclusion applies and we have checked that this remains valid even if we vary $\gamma$, as long as we take it to be of order one.

\subsubsection*{Marginalised constraints on $\Hend$}

\begin{figure}[t]
\begin{center}
\includegraphics[width=0.7\textwidth]{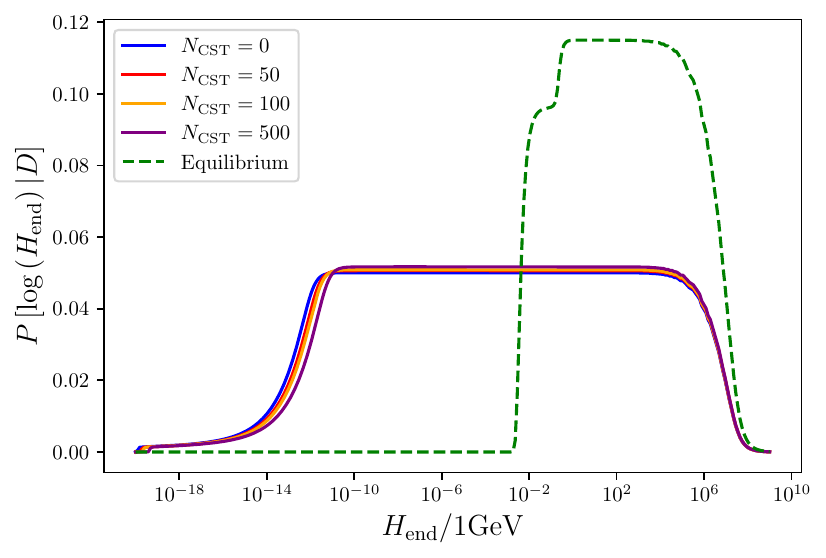}
\caption{Posterior distribution on $\log\Hend$, obtained from marginalising the full posterior distribution~\eqref{eq:posterior:logHend_logf}, displayed in \Fig{fig:EB-2d-1plot}, over $\log f$. Different colours correspond to different values of $\NCST$. The dashed curve corresponds to the equilibrium assumption (see bottom panel of \Fig{fig:EB-2d-1plot}), from which values of $\Hend \lesssim 10\, \MeV$ are strongly disfavoured.}
\label{fig:EB-marginalized-on-Hend}
\end{center}
\end{figure}

Assuming the equilibrium distribution at the end of inflation, we saw that $\Hend$ cannot be smaller than $10^{-3}\GeV$, otherwise the value for $f$ that is required to accommodate the observed abundance of dark matter is super-Planckian. However, when the stochastic dynamics of the axion during inflation is correctly accounted for, $\Hend$ can be as low as $10^{-20}\GeV$, which matches the bound imposed by the requirement that inflation proceeds before big-bang nucleosynthesis. This difference in the range of allowed values for $\Hend$ can be seen more clearly by marginalising \Eq{eq:posterior:Hend_f} over $f$,
\bea
P\left (\Hend \vert D\right) = \int P\left(\Hend,f\vert D\right)\dd f \, ,
\eea
or \Eq{eq:posterior:logHend_logf} over $\log f$,
\bea
P\left (\log\Hend \vert D\right) = \int P\left(\log\Hend,\log f\vert D\right)\dd (\log f) = \Hend\, P\left (\Hend \vert D\right) \, .
\eea

The result is displayed as a blue solid curve ($\NCST=0$) in \Fig{fig:EB-marginalized-on-Hend} where we can see that, contrary to what the equilibrium assumption suggests (see the dashed curve in \Fig{fig:EB-marginalized-on-Hend}), low-scale inflation (\ie small values of $ \Hend $) is allowed. At the $ 2\sigma $ level\footnote{Hereafter, when quoting $ 2\sigma $ constraints from the posterior distribution, the central value corresponds to the mean value, while the bounds are such that the probability for $\Hend$ to be smaller (larger) than the lower (upper) bound, is half of $4.55\%$. If the distribution were Gaussian, this would reduce to the usual $ 2 \sigma $ confidence interval.}, we find $ \log_{10}\left(\Hend\slash 1\GeV\right) = -3.4^{ +9.9}_{- 10.4}$ while under the equilibrium assumption we obtain $ \log_{10}\left(\Hend\slash 1\GeV\right) = 2.3^{ +4.6}_{- 4.4}$.

\subsubsection*{Marginalised constraints on $f$}

\begin{figure}[t]
\centering
\includegraphics[width=0.49\textwidth]{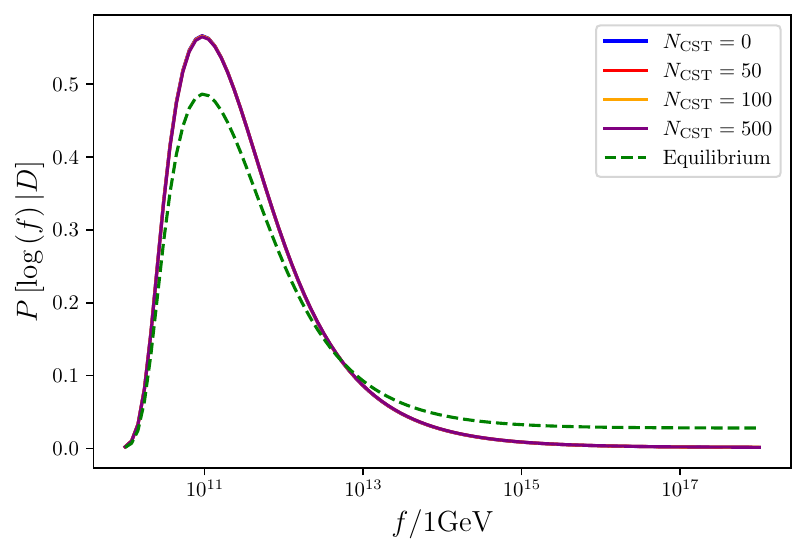}
\includegraphics[width=0.49\textwidth]{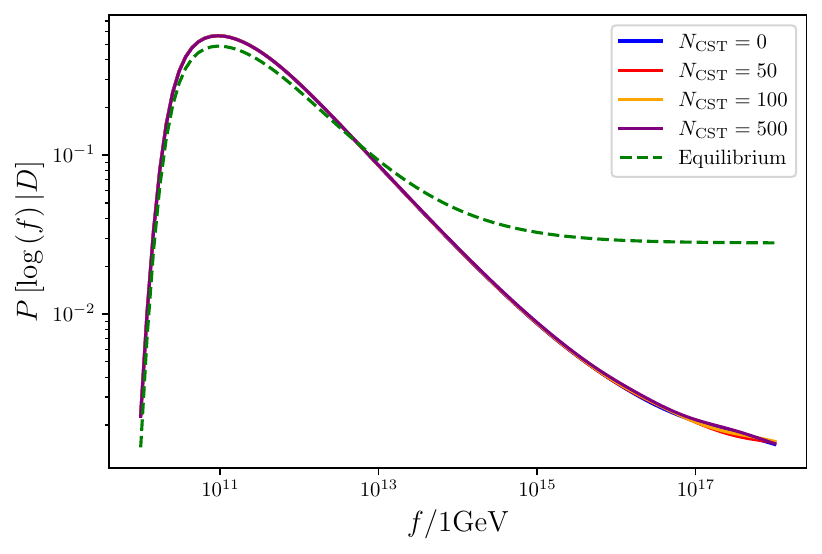}
\caption{Posterior distribution on $\log f$, obtained from marginalising the full posterior distribution~\eqref{eq:posterior:logHend_logf}, displayed in \Fig{fig:EB-2d-1plot}, over $\log\Hend$. Different colours correspond to different values of $\NCST$, which can hardly be distinguished. The dashed curve corresponds to the equilibrium assumption (see bottom panel of \Fig{fig:EB-2d-1plot}), from which large values of $f$ are less disfavoured. The two panels display the same curves but the right panel uses a logarithmic scale on the vertical axis to better stress the effect of the equilibrium assumption on the tail.}
\label{fig:EB-marginalized-on-f}
\end{figure}

Similarly, the marginalised distribution over $f$ can be considered, and is displayed in \Fig{fig:EB-marginalized-on-f}. Although the most favoured values of $f$ lie around $f\sim 10^{12}\GeV$ with or without the equilibrium assumption, one can see that, under the equilibrium assumption, a heavy tail is obtained, and large values of $f$ are also allowed, up to the Planck scale. However, when the stochastic dynamics of the axion is correctly accounted for during inflation, the distribution is strongly suppressed at large values. At the $ 2\sigma $ level, we find $ \log_{10}\left(f\slash 1\GeV\right) = 11.6^{ + 2.5}_{ - 1.3} $ while under the equilibrium assumption we obtain $ \log_{10}\left(f\slash 1\GeV\right) = 12.1^{+ 5.1}_{ - 1.7 }$.

\subsubsection*{Role of the plateau phase}

\begin{figure}[t]
\begin{center}
\includegraphics[width=1\textwidth]{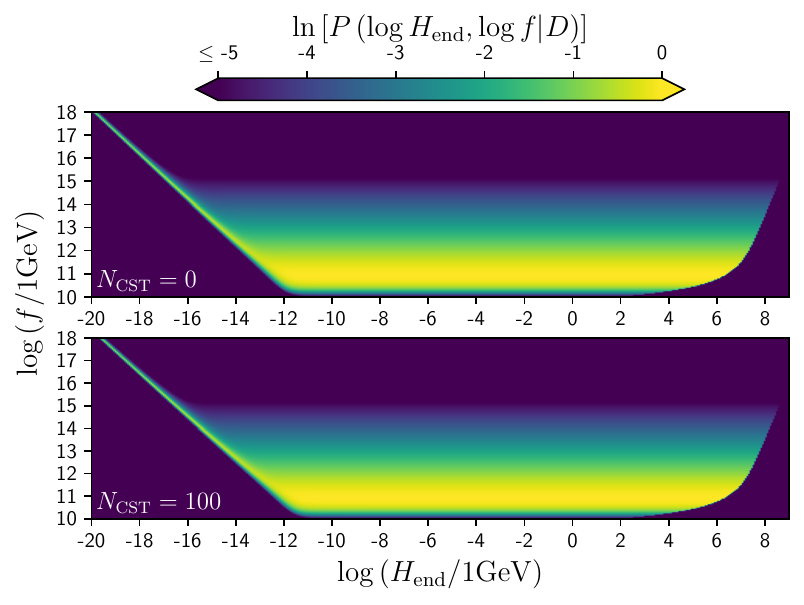}
\caption{Posterior distribution as a function of $\log f $ and $\log \Hend $, defined in \eqref{eq:posterior:logHend_logf}. The top panel is obtained with $ \NCST = 0 $ (hence it is the same as the top panel of \Fig{fig:EB-2d-1plot}), while the bottom panel corresponds to $ \NCST = 100 $. The comparison reveals a slight shift in the stochastic window when increasing $\NCST$ for $\Hend<10^{12}\GeV$. However, for larger values of $\Hend$, the distribution is almost unaffected.}  
\label{fig:EB_comparison}
\end{center}
\end{figure}

As discussed in \Sec{sec:plateau}, even though the bulk of the inflationary potential may be expected to be of the large-field type, from CMB observations it is known that the last $\sim 60$ \efolds of inflation at least proceed in a plateau potential. It is important to study how this ultimate plateau phase affects our predictions.
In \Fig{fig:EB_comparison} we display the posterior distribution~\eqref{eq:posterior:logHend_logf} for $ \NCST = 0 $ and $ 100 $ and observe a slight shift of the stochastic window when $ \NCST $ increases. The effect is tiny and only the tilted part of the axionic window is slightly shifted to larger values of $\Hend$. This is because, as explained above, in the majority of the axionic window the axion is very light at the end of inflation, hence its distribution does not vary substantially on time scales of order a few hundreds of \efolds. It does only if its mass is comparable to the Hubble expansion rate at the end of inflation, which corresponds to the tilted part of the axionic window.

\begin{figure}[t]
\begin{center}
\includegraphics[width=0.7\textwidth]{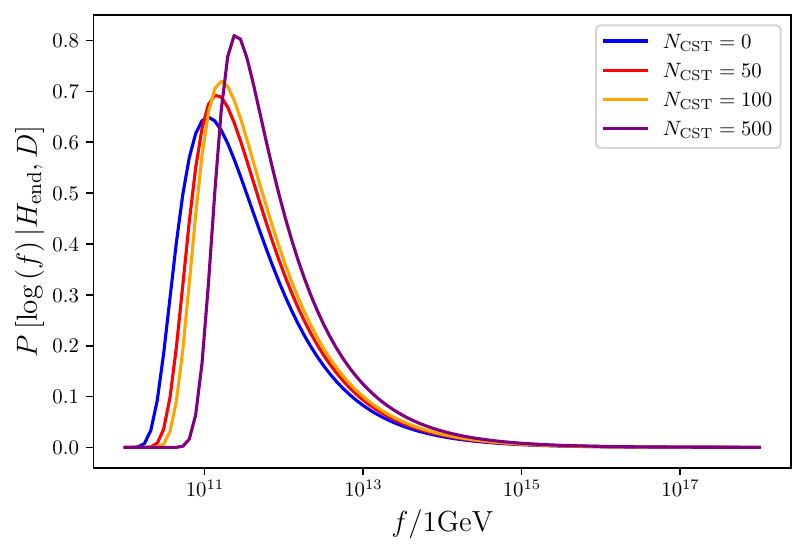}
\caption{Conditional posterior distribution $ P\left(\log f\vert\Hend,D\right) $ for a few values of $ \NCST $ for $ \Hend = 10^{-12}\, \GeV $.}  
\label{fig:slice-conditional-probability}
\end{center}
\end{figure}

This can be also seen in \Fig{fig:EB-marginalized-on-f}, where the marginalised distribution of $f$ is displayed for a few values of $\NCST$ up to $500$, and no effect from the plateau phase can be seen by eye. In \Fig{fig:slice-conditional-probability}, we display the conditional distribution of $\log f$ for a fixed value of $\Hend=10^{-12}\GeV$, which consists in setting $\Hend$ to that value in \Eq{eq:posterior:logHend_logf} and multiplying it with the $f$-independent normalization factor $1/P(\log\Hend)$. The effect of varying $\NCST$ is more pronounced in that case since the tilted part of the window is being probed. Adding a phase of plateau inflation lets the axion distribution narrow down at the end of inflation, which has to be compensated by larger values of $f$ to accommodate the observed dark-matter abundance. This explains why the conditional posterior distribution of $f$ shifts to slightly larger values when $\NCST$ increases, although the effect on the marginalised posterior distribution is negligible in \Fig{fig:EB-marginalized-on-f}.
 
 We thus conclude that our results are mostly unchanged by the addition of a phase of plateau inflation of a few hundreds of \efolds, which is otherwise needed to account for the CMB measurements.

\section{Conclusion}
\label{sec:Conclusion}

In this work, we have re-examined the so-called stochastic window for QCD axion dark matter. This corresponds to the values of $f$ (the axion decay constant) and $\Hend$ (the Hubble expansion rate at the end of inflation) for which the axion may constitute all of dark matter. The abundance of axion particles today is determined by the vacuum expectation value the axion field acquires during inflation. This, in turn, is driven by its classical rolling along the potential it receives below the QCD symmetry breaking scale $\Lambda\sim 100\MeV$, and by quantum diffusion, which emerges from the amplification of vacuum quantum fluctuations during inflation, and their stretching above the Hubble radius. This leads to a stochastic distribution for the axion misalignment angle at the end of inflation, on which the present paper has focused.

Measurements of the CMB favour slow-roll models of inflation, where $H$ varies slowly during inflation. If $H$ is constant, the axion is attracted towards an equilibrium distribution, which is flat if $H > \Lambda$ and peaked otherwise. This has been used in the literature to assess the  current axion abundance. In this setup, for the axion to constitute all of dark matter, one requires $10^{10.4}\GeV\leq f\leq 10^{17.2}\GeV$ and 
$\Hend>10^{-2.2}\GeV$, which is the conventional stochastic axion window.

However, we have shown that in practice, the equilibrium distribution is never reached at the end of inflation. In fact, reaching the equilibrium requires unrealistic single-field inflationary potentials (such as plateau phases lasting more than $10^{20\sim 30}$ \efolds without receiving any radiative corrections) and/or setting initial conditions in the eternal-inflation regime. This is because, although $H$ varies slowly during inflation, it nonetheless varies more quickly than the time it takes for the axion to reach the equilibrium, which is of order $H^2 f^2/\Lambda^4$ in number of \efolds for $H\ll \Lambda$ and $f^2/H^2$ for $H\gg \Lambda$. Although this does not preclude the existence of more involved inflationary setups where an equilibrium distribution is reached by the axion during inflation, see \eg \Refs{Matsui:2020wfx, Kadota:2003fs, Kadota:2003tn, Kitajima:2019ibn}, this implies that in the single-field models currently preferred by cosmological measurements, the stochastic dynamics of the axion during inflation needs to be re-examined. In particular the axion distribution at the end of inflation is sensitive to the unknown initial distribution unless some assumptions are made for the large field region of the inflaton potential.

When the ubiquitous monomial corrections are included in the large-field regions of the inflationary potential, we have found that the misalignment angle has a flat distribution at the end of inflation as long as $\Hend>10^{-13.8}\GeV$, \ie as long as the axion remains light during inflation. We have checked that our results are robust under considering other types of large-field corrections to the inflationary potential, and that they are unaffected by the phase of plateau inflation that needs to take place in the last hundreds of \efolds of inflation to account for CMB observations. We have also shown that our conclusions are independent of initial conditions, as long as they are set above a critical scale that we have determined, see \Eq{eq:Hc:gen}. 

For smaller values of $\Hend$, the distribution becomes more peaked, which allows values of $f$ larger than $10^{12}\GeV$ to account for the observed dark matter abundance. However, when parameter constraints are properly reconstructed in a Bayesian framework (to account for the stochastic nature of the misalignment angle), we find that this requires a substantial amount of fine tuning, unless a fundamental theory somehow predicts a prior distribution $P(\Hend,f)$ sharply peaked around $f\sim \Lambda^2/\Hend$, which at least requires a close connection between two different physics (inflation and a spectator axion field). Thus under the assumption of flat prior on the log scale for $\Hend$ and $f$ we show that 
\bea
\label{eqn:mainconclusion}
10^{10.3}\GeV\leq f\leq 10^{14.1}\GeV\, , \quad\Hend>10^{-13.8}\GeV
\eea 
is the correct stochastic window. This is the main conclusion of this work, which has important consequences for axion phenomenology in Cosmology. Indeed, $f\lesssim 10^{13.4}\GeV$ implies that the mass of the QCD axion is of the order $m\gtrsim 10^{-4}$eV. Such axion mass is above the mass range which the current microwave cavity experiments are sensitive to 
\cite{ParticleDataGroup:2022pth,AxionLimits}.  Moreover, this implies that there might be less need to significantly increase the size of cavities in microwave cavity experiments, contrary to what the equilibrium assumption suggests when allowing for smaller axion masses. The result \eqref{eqn:mainconclusion} also implies that low-scale inflation is fully compatible with QCD axion dark matter, contrary to previous claims.

\acknowledgments
We thank the Yukawa Institute for Theoretical Physics at Kyoto University, where 
this work was initiated during the stay under ``Visitors Program of FY2022''. KK was partly supported by Center of Quantum Cosmo Theoretical Physics (NSFC grant number 12347103) and thanks the CTPU at IBS for hospitality during the completion of this work. 
The work of SM was supported in part by the World Premier International Research Center Initiative (WPI), MEXT, Japan. SM is grateful for the hospitality of Perimeter Institute, the cosmology group at Simon Fraser University and the Theoretical Physics Institute at University of Alberta, where part of this work was carried out. Research at Perimeter Institute is supported in part by the Government of Canada through the Department of Innovation, Science and Economic Development and by the Province of Ontario through the Ministry of Colleges and Universities.
 AT would like to thank University of Rwanda, EAIFR, and ICTP for their kind hospitalities when some parts of the project were in hand.

\appendix

\section{Criterion for oscillations}
\label{app:Oscillations}

In this appendix, we derive a criterion for the occurrence of axion oscillations during inflation. When $m\ll \Hend$, the axion remains light throughout inflation and evolves in the slow-roll regime, where oscillations do not take place. When $\Hend$ decreases and become of order $m$, the axion may oscillate if its initial value is small enough. Given that oscillations first occur for very small values of $\theta$, to derive the critical value of $\Hend$ one can linearise the Klein-Gordon equation~\eqref{eq:KG}, which leads to
\begin{equation}
\label{Klein-Gordon linearized}
    \frac{\dd^2\theta}{\dd N^2} + (3-\epsilon_1)\frac{\dd\theta}{\dd N} + \frac{m^2}{H^2}\theta = 0\, .
\end{equation}
The friction term can be removed under introducing $ \theta(N) = g(N) y(N) $, where $g$ satisfies the first-order differential equation
\bea
\label{eq:ode:g}
\frac{\dd g}{\dd N} + \frac{3-\epsilon_1}{2}g = 0\, .
\eea
This can be solved as $g\propto \ee^{-\frac{3}{2}N+\frac{1}{2}\int^N\epsilon_1(N')\dd N'}$, which reduces to
\bea
\label{eq:g:sol}
g = g_\uend \ee^{-\frac{3}{2}\left(N-N_\uend\right)} \left[1+2\left(N_\uend-N\right)\right]^{-1/4}
\eea
in quadratic inflation. With this change of variables one obtains
\begin{equation}
    \frac{\dd^2y}{\dd N^2} + \omega^2y = 0,
\end{equation}
where $\omega^2=\epsilon_1 \epsilon_2/2 - (3-\epsilon_1)^2/4+m^2/H^2$.
Here, $\epsilon_2=\dd\ln\epsilon_1/\dd N$ is the second Hubble-flow parameter, and in quadratic inflation it is given by $\epsilon_2=2\epsilon_1$. One thus obtains a Whittaker equation,
\bea
    \frac{\dd ^2y}{\dd v^2} + \left(-\frac{1}{4} + \frac{\kappa}{v} + \frac{\frac{1}{4}-\mu^2}{v^2}\right)y = 0,
\eea
where $ v = \frac{3}{2}[1+2(N_{\mathrm{end}}-N)] $, $ \mu = \frac{1}{4} $ and $ \kappa = \frac{1}{4} + \frac{m^2}{6\Hend^2} $. It can be solved as
\bea
y(v) = A W_{\kappa,\mu}(v) + B M_{\kappa,\mu}(v)\, ,
\eea
where $W_{\kappa,\mu}$ and $M_{\kappa,\mu}$ are the two Whittaker functions. In the asymptotic past, $v$ is large, and one has $W_{\kappa,\mu}(v)\simeq \ee^{-v/2}v^\kappa$ and $M_{\kappa,\mu}(v)\simeq \ee^{v/2}v^{-\kappa}\Gamma(2\mu+1)/\Gamma(\mu-\kappa+1/2)$ when $v\gg 1$. In this limit, $v\simeq 3(N_\uend-N)$, and together with \Eq{eq:g:sol} one finds
\bea
\label{eq:theta:sol:quad:past}
\theta \simeq \bar{A}\left[1+2\left(N_\uend-N\right)\right]^{\kappa-\frac{1}{4}} + \bar{B}\ee^{3\left(N_\uend-N\right)}\left[1+2\left(N_\uend-N\right)\right]^{-\kappa-\frac{1}{4}}\, ,
\eea
where $\bar{A}$ and $\bar{B}$ are rescaled versions of $A$ and $B$, namely $\bar{A}=A g_\uend (3/2)^\kappa \ee^{-3/4}$ and $\bar{B}=B g_\uend (3/2)^{-\kappa} \ee^{3/4}\Gamma(2\mu+1)/\Gamma(\mu-\kappa+1/2)$. Let us now compare this behaviour with the slow-roll solution, towards which the axion is attracted at early time, and which corresponds to solving \Eq{Klein-Gordon linearized} in the absence of the acceleration term $\dd^2\theta/\dd N^2$. One obtains
\bea
\theta \propto \exp\left\lbrace - \int^N \frac{m^2}{H^2(\tilde{N})\left[3-\epsilon_1(\tilde{N})\right]} \dd \tilde{N}\right\rbrace  \, .
\eea
Note that $\epsilon_1$ quantifies the slow rolling of the inflaton, not the axion, which is why it is kept in the above, although its contribution in the asymptotic past is negligible. In quadratic inflation, this leads to
\bea
\theta\propto \left[1+\left(N_\uend-N\right)\right]^{\frac{m^2}{6\Hend^2}}\, .
\eea 
This precisely coincides with the $\bar{A}$ branch of \Eq{eq:theta:sol:quad:past}, which can thus be interpreted as the slow-roll attractor. The fact that the non-attractive branch (here the $\bar{B}$ branch) decays as $\ee^{-3N}$ is in fact expected in slow-roll models, see for instance \Refs{Vennin:2014xta, Chowdhury:2019otk}. Requiring initial conditions to be along the slow-roll attractor thus amounts to setting $B=0$, hence 
\bea
\theta(v) \propto \ee^{-\frac{v}{2}} v^{-1/4} W_{\kappa,\mu}(v)\, ,
\eea 
where the overall constant is set by the initial condition $\theta(v_\uin)=\theta_\uin$.

At the end of inflation, $ v = \frac{3}{2} $ so the largest value of $\Hend$ below which oscillations occur is such that $ W_{\kappa,\mu}(\frac{3}{2}) = 0 $. With $\mu=1/4$, this leads to $ \kappa = 7/4 $, hence $\Hend = m/3$. We conclude that oscillations of the axion take place before inflation ends if and only if
\bea
m>3\Hend\, .
\eea 
Note that this result for quadratic inflation is exact and does not rely on approximations.

\bibliographystyle{JHEP}
\bibliography{./ref}

\end{document}